\documentclass[preprint,
amssymb,amsmath,aps,floatfix,nofootinbib]{revtex4-1}
  \usepackage{latexsym,bm,amsmath,amssymb,amsfonts}
\newcommand{\beq}{\begin{eqnarray}}
\newcommand{\eeq}{\end{eqnarray}}

\newcommand{\nn}{\nonumber \\}

\usepackage{graphicx}
\usepackage{amsmath,amsthm,amssymb}

\begin{document}

\preprint{YITP-14-21}

\title{A systematic study of exact solutions in second-order conformal hydrodynamics}
\author{Yoshitaka Hatta}
\affiliation{Yukawa Institute for Theoretical Physics, Kyoto University, Kyoto 606-8502, Japan}
\author{Jorge Noronha}
\affiliation{Instituto de F\'isica, Universidade de S\~ao Paulo, C.P. 66318, 05315-970 S\~ao Paulo, SP, Brazil}
\author{Bo-Wen Xiao}
\affiliation{Key Laboratory of Quark and Lepton Physics (MOE) and Institute
of Particle Physics, Central China Normal University, Wuhan 430079, China}

\date{\today}
\vspace{0.5in}
\begin{abstract}
In this paper we present the details of our previous work on exact solutions in second-order conformal hydrodynamics together with a number of new solutions found by mapping Minkowski space onto various curved spacetimes such as anti-de Sitter space and hyperbolic space. We analytically show how the solutions of ideal hydrodynamics are modified by the second-order effects including vorticity. We also find novel boost-invariant exact solutions which exist only in second-order hydrodynamics and have an unusual dependence on the proper time.
\end{abstract}

\maketitle

\section{Introduction}
Relativistic hydrodynamics is a ubiquitous tool to address long-wavelength phenomena in various areas of high energy physics such as astrophysics and
heavy-ion physics which reveals physical properties of the Quark-Gluon Plasma (QGP). 
In practical applications, hydrodynamics boils down to a set of  nonlinear coupled partial differential equations of many variables which are almost always solved numerically. However, under sufficient symmetry conditions, it is often possible to derive exact analytical solutions. Classic examples are the Hubble flow in cosmology \cite{weinberg} and the Bjorken flow \cite{Bjorken:1982qr} in relativistic heavy-ion collisions. Many other solutions of ideal hydrodynamics have been found in the  literature,  mostly in the context of studies of QGP/heavy-ion physics \cite{Biro:2000nj,Csorgo:2003rt,Csorgo:2003ry,Csorgo:2006ax,Friess:2006kw,Bialas:2007iu,Beuf:2008vd,peschanskiother,
 Fouxon:2008ik,Nagy:2009eq,Liao:2009zg,Lin:2009kv}.
These analytical solutions provide us with good physical intuition into the problem and they can also serve as a test of numerical hydrodynamic codes.

On the other hand, attempts to analytically solve \emph{non}-ideal relativistic hydrodynamic equations have been scarce, if not nonexistent. Even the Navier-Stokes equation, which includes only the first-order viscous corrections to ideal hydrodynamics, is significantly more complicated to tackle analytically beyond perturbation theory. Moreover, finding solutions of the relativistic Navier-Stokes equation is not entirely satisfactory because, as is well-known, the equation has serious drawbacks which are only remedied by including second-order corrections. The precise complete formulation of second-order hydrodynamics is still under active debate, but it typically contains ${\mathcal O}(10)$ new terms and transport coefficients, making it a daunting task to obtain any analytical insights. This seems a bit frustrating in view of the recent progress in the foundation of second-order relativistic hydrodynamics \cite{DeGroot:1980dk,Koide:2006ef,Muronga:2006zx,Baier:2007ix,Bhattacharyya:2008jc,Betz:2008me,
PeraltaRamos:2009kg,Denicol:2011fa,Denicol:2012cn,Tsumura:2013uma}, and all the more so because the experimental data from heavy-ion collisions have shown indications of non-ideal fluid behavior and, thus, viscous hydrodynamics simulations are increasingly becoming a standard tool to analyze the data \cite{Heinz:2013th}.

\emph{Conformal symmetry} offers powerful methods to solve difficult problems in field theory which are otherwise intractable \cite{yellowpages}, and here again, it proves to be useful.  The hydrodynamic equations are greatly simplified in the presence of conformal symmetry not only because it puts constraints on various second-order terms \cite{Baier:2007ix,Bhattacharyya:2008jc}, but also because it allows us to use the Weyl rescaling of the metric
\beq
g_{\mu\nu} \to \Lambda^2 \hat{g}_{\mu\nu}\, , \label{wey}
\eeq
where $\Lambda$ is an arbitrary scalar function of the coordinates, to work in a convenient space-time where the problem simplifies. This latter attribute has been recently exploited in \cite{Gubser:2010ze,Gubser:2010ui} to find an exact solution of the relativistic Navier-Stokes equation. This approach was further extended in \cite{Marrochio:2013wla} where semi-analytical solutions (as well as an approximate analytic solution) of Israel-Stewart theory \cite{Israel:1979wp} were found. 

In a previous paper \cite{Hatta:2014gqa}, by using the Weyl equivalence between Minkowski space and $AdS_3\times S^1$, we have constructed some exact solutions in second-order conformal hydrodynamics which are valid for rather generic values of the transport coefficients involved. In this paper we present the details of this work and derive a number of novel second-order solutions by conformally mapping Minkowski space to various space-times (hyperbolic space, anti-de Sitter space, etc.).  We also find new boost-invariant conformal fluid solutions which are similar to the Bjorken solution but possess an unconventional (though natural from the point of view of conformal invariance) time-dependence.

This paper is organized as follows. After introducing the basics of second-order hydrodynamics in Section \ref{secondorder}, we describe various exact solutions of ideal conformal hydrodynamics in Section \ref{solution}. We then include the second-order corrections to some of these solutions and construct new solutions  in the irrotational case (Section \ref{seco}) and in the rotating case (Section \ref{rotation}). In Section \ref{shear} we revisit the boost-invariant problem and find special solutions for the most general conformal second-order equation. Section \ref{conc} is devoted to conclusions.
\section{Second-order hydrodynamic equations}
\label{secondorder}

In this section we review the second-order formalism of relativistic hydrodynamics and set up our notations.
The energy-momentum tensor of a relativistic fluid is parametrized in the usual way \cite{DeGroot:1980dk}
\beq
T^{\mu\nu}=\epsilon u^\mu u^\nu + (p+\Pi)\Delta^{\mu\nu} + \pi^{\mu\nu}\,.
\eeq
$\epsilon$ is the energy density and $p$ is the (thermodynamic) pressure.
$u^\mu$ is the flow velocity normalized as $u^\mu u_\mu=-1$  and $\Delta^{\mu\nu}=g^{\mu\nu}+u^\mu u^\nu$ is the projection operator transverse to the flow with $g_{\mu\nu}=(-,+,+,+)$. The bulk pressure $\Pi$ and the shear-stress tensor $\pi^{\mu\nu}$ characterize the deviation from local equilibrium. We work in the so-called Landau frame \cite{DeGroot:1980dk} in which $\pi^{\mu\nu}$ is transverse $u_\mu \pi^{\mu\nu}=0$ and traceless $\pi^\mu_{\ \mu}=0$.

Throughout this paper, we assume that there are no other macroscopic conserved currents besides energy and momentum. Therefore, there are 11 unknown variables $\epsilon, p, u^\mu, \Pi, \pi^{\mu\nu}$ which should be  determined by 11 equations. In the presence of  conformal symmetry, this number becomes 9 because $\Pi=0$ due to the traceless condition $T^\mu_{\ \mu}=0$ and  $\epsilon$ and $p$ are related by the equation of state
\beq
p=\frac{1}{3}\epsilon\,.
\eeq

Four equations are provided by the
energy-momentum conservation law $\nabla_\mu T^{\mu\nu}=0$ ($\nabla_\mu$ is the space-time covariant derivative). This can be decomposed into the components parallel and transverse to the flow as
\beq
&& D \epsilon + (\epsilon +p )\vartheta+\pi^{\mu\nu}\sigma_{\mu\nu}=0\,, \label{a} \\
&& (\epsilon +p ) D u^\mu + \Delta^{\mu\alpha}\nabla_\alpha p  + \Delta^\mu_{\ \nu}\nabla_\alpha \pi^{\alpha\nu}=0\,, \label{b}
\eeq
 where we already set $\Pi=0$ and defined the comoving derivative $D\equiv u^\mu \nabla_\mu$.
$\vartheta\equiv \nabla_\mu u^\mu$ is the fluid expansion rate and
\beq
\sigma^{\mu\nu}\equiv \nabla^{\langle \mu} u^{\nu\rangle} \equiv  \left(\frac{1}{2}(\Delta^{\mu\alpha}\Delta^{\nu\beta}
+\Delta^{\mu\beta}\Delta^{\nu\alpha})-\frac{1}{3}\Delta^{\mu\nu}\Delta^{\alpha\beta} \right)\nabla_\alpha u_\beta\,,
\eeq
 is the shear tensor. The brackets on Greek indices $A^{\langle \mu\nu\rangle}$ denote the projection onto the transverse and traceless part of the tensor $A^{\mu\nu}$.

The remaining five equations for the five components of $\pi^{\mu\nu}$  describe the space-time dependence of these dissipative currents. Since the work of Israel and Stewart \cite{Israel:1979wp}, there has been a longstanding controversy regarding the precise structure of these equations in relativistic systems \cite{DeGroot:1980dk,Muronga:2006zx,Baier:2007ix,Bhattacharyya:2008jc,Koide:2006ef,Betz:2008me,
PeraltaRamos:2009kg,Denicol:2011fa,Denicol:2012cn,Tsumura:2013uma}. Here we employ  the result of  Denicol \emph{et al.} \cite{Denicol:2012cn} and generalize it to curved spacetimes taking into account the constraints from conformal symmetry \cite{Baier:2007ix,Bhattacharyya:2008jc}. The most general equation then reads\footnote{The equation  derived in \cite{Denicol:2012cn} includes terms proportional to the pressure gradient $F^\mu=\Delta^{\mu\nu}\nabla_\nu p$ (or equivalently, the temperature gradient). Eliminating them by using Eq.\ (\ref{b}) gives rise to a new term $D\sigma^{\mu\nu}$ and modifies the coefficient of other terms accordingly \cite{Baier:2007ix}. Thus the various transport coefficients shown in (\ref{d}) are in general different from the corresponding ones in Ref.~\cite{Denicol:2012cn}. }
\beq
\pi^{\mu\nu} &=&-2\eta \sigma^{\mu\nu} - \tau_\pi \left(\Delta^\mu_\alpha \Delta^\nu_\beta  D\pi^{\alpha\beta}   +\frac{4}{3}  \pi^{\mu\nu} \vartheta \right) +\lambda_2 \pi^{\langle \mu}_{\ \ \lambda} \Omega^{\nu\rangle\lambda} \nn
&& + \lambda_1 \pi^{\langle \mu}_{\ \ \lambda}\pi^{\nu\rangle\lambda}+ \lambda_3 \Omega^{\langle \mu}_{\ \  \lambda}\Omega^{\nu\rangle \lambda}
-\tau_{\pi\pi} \sigma^{\langle \mu}_{\ \  \lambda}\pi^{\nu\rangle \lambda}
- \tilde{\eta}_3\sigma^{\langle \mu}_{\ \  \lambda}\sigma^{\nu\rangle \lambda} - \tilde{\eta}_4 \sigma^{\langle \mu}_{\ \  \lambda}\Omega^{\nu\rangle \lambda} \nn
&&+\tau_\sigma\left( \Delta^\mu_\alpha \Delta^\nu_\beta  D \sigma^{\alpha\beta}+\frac{1}{3}\sigma^{\mu\nu}\vartheta \right) +\kappa \left(\mathcal{R}^{\langle \mu\nu\rangle} -2u_\alpha \mathcal{R}^{\alpha\langle\mu\nu\rangle\beta} u_\beta \right)
\,, \label{d}
\eeq
 where $\Omega^{\mu\nu}\equiv \frac{1}{2}\Delta^{\mu\alpha}\Delta^{\nu\beta}(\nabla_\alpha u_\beta -\nabla_\beta u_\alpha)$ is the vorticity tensor.
$\eta$ is the shear viscosity that appears in the first-order (Navier-Stokes) theory. In kinetic theory approaches valid at weak coupling, one typically finds the relation $\tau_\pi=2\lambda_2$ \cite{Israel:1979wp}.
 The linear combinations inside the brackets are designed to transform homogeneously under the Weyl transformation (\ref{wey}).
The last terms involving the Riemann and Ricci tensors are relevant to the dynamics only in curved spacetimes. In this paper, we do consider hydrodynamics in curved spacetimes but they are all conformally equivalent to flat Minkowski space. In this case the linear combination proportional to $\kappa$ vanishes identically and, therefore, it will not be  considered in the following.

In Ref.~\cite{Denicol:2012cn}, without assuming conformal symmetry, Denicol \emph{et al}. derived  the above equation for $\pi^{\mu\nu}$  in flat space-time via a consistent truncation of the Boltzmann equation doubly expanded in powers of the Knudsen number (expansion in the number of space-time gradients) and the inverse Reynolds number (expansion in the ratios of dissipative to equilibrium quantities) up to second order. Their method may be viewed as a relativistic generalization of Grad's moment method \cite{grad}, but unlike Grad's original theory or Israel-Stewart's relativistic theory containing only the first line of Eq.~(\ref{d}), it features a well-defined power counting scheme which allows one to systematically improve the approximation involved.

An important concept underlying this (generalized) moment method  is that  $\pi^{\mu\nu}$ should be treated as \emph{independent} variables which are determined self-consistently and non-linearly from Eq.\ (\ref{d}). This is actually crucial to our work. In the literature, one often treats  $\pi^{\mu\nu}$ and $-2\eta \sigma^{\mu\nu}$ interchangeably in the second-order terms \cite{Baier:2007ix}.  Then there is no longer any essential distinction between $\pi\pi$, $\sigma\sigma$ and $\pi\sigma$ terms, or $D\sigma$ and $D\pi$ terms so that Eq.~(\ref{d}) reduces to a gradient expansion. While such an identification may be justified for certain purposes, such as finding perturbative asymptotic solutions, throughout this paper we shall encounter examples in which the approximation $\pi^{\mu\nu}\approx -2\eta \sigma^{\mu\nu}$ is violated or makes no sense.
This is most obvious when $\sigma^{\mu\nu}=0$ (note that Eq.~(\ref{d}) is nontrivial and well-defined even in this case), and in Section \ref{shear} we shall see an explicit violation of this approximation when $\sigma^{\mu\nu}\neq 0$.\footnote{The analytical solution obtained in \cite{Marrochio:2013wla} also violates this approximation.} More generally, Eq.~(\ref{d}) and its variants based on the gradient expansion admit qualitatively different solutions. The latter have long been known to be unstable in nonrelativistic theory \cite{bob},
and acausal \cite{acausal} in the relativistic domain 
(see also \cite{Pu:2009fj,Minami:2013aia} for  recent discussions). Because of this, the relativistic hydrodynamic equations obtained from the gradient expansion are not usually implemented in numerical hydrodynamic studies.

\section{Solutions of conformal ideal hydrodynamics}
\label{solution}

In this section we describe various exact solutions of the ideal hydrodynamic equations, namely, (\ref{a}) and (\ref{b}) with $\pi^{\mu\nu}=0$. Some of the results in this section are new. All the solutions  are obtained by the following general strategy: we first consider a coordinate transformation from the Minkowski coordinates $(t,\vec{r})$ to some curvilinear coordinates $x^\mu=x^\mu(t,\vec{r})$
\beq
ds^2=-dt^2+dx^2+dy^2+dz^2= g_{\mu\nu}dx^\mu dx^\nu\,.
\eeq
If there is conformal symmetry, this may be combined with the Weyl rescaling of the metric \cite{Gubser:2010ze}
\beq
ds^2  = \Lambda^2 \hat{g}_{\mu\nu} d\hat{x}^\mu d\hat{x}^\nu \equiv \Lambda^2 d\hat{s}^2\,. \label{conf}
\eeq
We then identify the the static, or comoving solution with respect to the new `time' coordinate $x^0$ or $\hat{x}^0$. When transformed back to Minkowski space, this becomes a nontrivial solution which depends on the original space-time coordinates ($t, \vec{r}$).

  Throughout this paper, we  use the notation $r=\sqrt{x^2+y^2+z^2}$, $x_\perp=\sqrt{x^2+y^2}$. The spherical coordinates are defined as usual, $\cos \theta=\frac{z}{r}$ and $\tan\phi=\frac{y}{x}$, with the line element $d\Omega^2=d\theta^2+\sin^2\theta d\phi^2$.
 Also, as already done in (\ref{conf}), we  use a `hat' (e.g., $\hat{\epsilon}$) for quantities in the Weyl-transformed coordinates.

\subsection{Bjorken flow }
\label{bjo}

Bjorken's solution \cite{Bjorken:1982qr} provides a useful approximation of the complicated dynamics of the matter created in ultrarelativistic heavy-ion collisions. The flow expands in the beam direction (along the $z$-axis), and is naturally described in the coordinate system
\beq
ds^2=-d\tau^2+ \tau^2d\eta^2 +dx^2+dy^2\,,
\eeq
 where  $\tau\equiv \sqrt{t^2-z^2}$ is the `proper time' and $\eta \equiv \tanh^{-1}\frac{z}{t}$ is the `rapidity' (we use the same letter $\eta$ for the rapidity and the shear viscosity but the distinction should be obvious from the context).

In this coordinate system, the comoving solution is characterized by the flow velocity
\beq
 u_\tau=-1\,, \qquad u_\eta=u_x=u_y=0\,. \label{como}
 \eeq
  This is a boost-invariant (i.e., independent of $\eta$) flow which  has an infinite extent in the transverse $(x,y)$ directions. In the original coordinates the flow velocity becomes
\beq
u^\mu=\left(\frac{t}{\tau},0,0, \frac{z}{\tau}\right)\,. \label{bj}
\eeq
The expansion parameter, the shear tensor, and the vorticity tensor are readily calculated as
\beq
\vartheta = \frac{1}{\tau}\,, \qquad \sigma^{\eta}_{\ \eta}=\frac{2}{3\tau}\,, \qquad \sigma^{x}_{\ x}=\sigma^{y}_{\ y}=-\frac{1}{3\tau}\,, \qquad \Omega^{\mu\nu}=0\,. \label{compo}
\eeq

The energy density can be determined by  the continuity equations (\ref{a}) and (\ref{b}).
The solution is well-known
\beq
\epsilon \propto \frac{1}{\tau^{4/3}}\,.  \label{bjorken}
\eeq
We note that for more general equations of state of the form $p=w\epsilon$, we have $\epsilon \propto 1/\tau^{1+w}$.

\subsection{Gubser flow}
\label{gubser}

Gubser generalized Bjorken's solution by including a nontrivial $x_\perp$-dependence while retaining boost invariance \cite{Gubser:2010ze,Gubser:2010ui}. Assuming conformal symmetry, Refs.~\cite{Gubser:2010ze,Gubser:2010ui} considered the following coordinate and Weyl  transformations of the metric
\beq
d\hat{s}^2\equiv \frac{ds^2}{\tau^2}
&=& \frac{-d\tau^2 + dx_\perp^2 + x_\perp^2 d\phi^2 }{\tau^2}+d\eta^2 \nn
&=& -d\varrho^2+ \cosh^2\varrho (d\Theta^2+\sin^2\Theta d\phi^2)+ d\eta^2\,, \label{ds}
\eeq
 where
 \beq
 \sinh\varrho = -\frac{L^2-\tau^2+x_\perp^2}{2L\tau}\,, \qquad \tan\Theta = \frac{2Lx_\perp}{L^2+\tau^2-x_\perp^2}\,. \label{gub}
 \eeq
 Eq.\ (\ref{ds}) shows that Minkowski space is conformal to $dS_3\times {\mathbb R}$ ($dS_3$ is the three-dimensional de Sitter space) up to a Weyl rescaling factor $\Lambda^2=\tau^2$. The parameter $L$ has the dimension of length and is identified with the `radius' of $dS_3$ (alternatively, \cite{Gubser:2010ze} defined an energy scale $q\equiv 1/L$)
 \beq
 -X_0^2+X_1^2+X_2^2+X_3^2=L^2\,,
 \eeq
 where
 \beq
 X_0=-\frac{L^2-\tau^2+x_\perp^2}{2\tau}\,, \qquad  X_{1,2}=\frac{x_\perp^{1,2}}{\tau}L\,, \qquad  X_3=\frac{L^2+\tau^2-x_\perp^2}{2\tau}\,.
 \eeq
Similarly to (\ref{como}), we choose the flow velocity as
\beq
\hat{u}_\varrho =-1, \qquad \hat{u}_\eta=\hat{u}_\Theta=\hat{u}_\phi=0\,. \label{gflow}
\eeq
This has the following properties
\beq
\hat{\vartheta}=2\tanh\varrho\,, \qquad \hat{\sigma}^{\eta}_{\ \eta}=-\frac{2}{3}\tanh\varrho\,,  \qquad \hat{\sigma}^\Theta_{\ \Theta}=\hat{\sigma}^{\phi}_{\ \phi}=\frac{1}{3}\tanh\varrho\,, \qquad \hat{\Omega}^{\mu\nu}=0
\,. \label{mori}
\eeq
 In the $\hat{x}$-coordinates, the continuity equations (\ref{a}) and (\ref{b}) take the form
 \beq
 3\hat{u}^\mu \hat{\nabla}_\mu\hat{\epsilon} + 4\hat{\epsilon}\, \hat{\vartheta}=0\,, \\
 4\hat{\epsilon}\hat{u}^\nu \hat{\nabla}_\nu \hat{u}^\mu +\hat{\Delta}^{\mu\alpha}\hat{\nabla}_\alpha \hat{\epsilon}=0\,. \label{ide}
 \eeq
Substituting $\vartheta$ from (\ref{mori}), we can easily find the solution
 \beq
 \hat{\epsilon} \propto \left(\frac{1}{\cosh\varrho}\right)^{\frac{8}{3}}\,.
 \eeq
The solution in Minkowski space is recovered by the coordinate and Weyl transformations
\beq
u_\mu &=& \Lambda \frac{\partial \hat{x}^\nu}{\partial x^\mu} \hat{u}_\nu\,, \\
\sigma_{\mu\nu} &=& \Lambda \frac{\partial \hat{x}^\alpha}{\partial x^\mu} \frac{\partial \hat{x}^\beta}{\partial x^\nu}\hat{\sigma}_{\alpha\beta}\,, \\
\epsilon &=& \frac{1}{\Lambda^4}\hat{\epsilon}\,,
\eeq
 where  the  power of $\Lambda=\tau$ indicates the conformal weight of the corresponding quantity \cite{yellowpages}.
In particular, the flow four-vector is given by
\beq
u_\tau=-\cosh\left[\tanh^{-1}\frac{2\tau x_\perp}{L^2+\tau^2+x_\perp^2}\right]\,,
\qquad
u_\perp = \sinh\left[\tanh^{-1}\frac{2\tau x_\perp}{L^2+\tau^2+x_\perp^2}\right]\,,
\label{tran}
\eeq
with $u_\eta=u_\phi=0$, and the energy density is
\beq
\epsilon \propto \frac{1}{\tau^{4/3}}\frac{1}{(L^4+2(\tau^2+x_\perp^2)L^2+(\tau^2-x_\perp^2)^2)^{4/3}}\,.
\eeq
In contrast to Bjorken's solution, the flow is expanding in the transverse direction with the velocity
\beq
v_\perp \equiv -\frac{u_\perp}{u_\tau}= \frac{2\tau x_\perp}{L^2+\tau^2+x_\perp^2}\,,
\eeq
which is reminiscent of the `radial flow' observed in actual heavy-ion experiments \cite{Heinz:2013th}. See \cite{Gubser:2010ui} for phenomenological applications of this solution to heavy-ion physics.

\subsection{Hubble flow in Milne universe}
\label{hubble}

The next example is the flat-space analog of the well-known Hubble flow in cosmology \cite{Csorgo:2003ry,Csorgo:2006ax}.
Let us consider the following transformation
\beq
ds^2 &=& -dt^2+dr^2+r^2d\Omega^2 \nn
&=&-d\tau_r^2+\tau_r^2d\eta_r^2 + \tau_r^2\sinh^2\eta_r d\Omega^2  \nn
&=& \tau_r^2 (-d\chi^2 + d\eta_r^2 + \sinh^2\eta_r d\Omega^2)\,, \label{we}
\eeq
where we defined $\tau_r\equiv \sqrt{t^2-r^2} \equiv e^\chi$ and $\eta_r\equiv \tanh^{-1}\frac{r}{t}$ and $d\Omega$ denotes the solid angle. These are the three-dimensional counterparts of the proper time $\tau=\sqrt{t^2-z^2}$ and rapidity $\eta=\tanh^{-1}\frac{z}{t}$ introduced in the previous subsections.

The second equality of Eq.~(\ref{we}) shows that the metric of the Milne universe is a special solution of the Friedmann equation \cite{weinberg} describing an empty universe with negative spatial curvature. As is well-known, and as is manifest in Eq.~(\ref{we}), the Milne universe is a simple reparametrization of Minkowski space. The third equality shows that Minkowski space is conformal to ${\mathbb R}\times H_3$ where $H_3$ is the three-dimensional hyperbolic space
\beq
X_0^2-X_1^2-X_2^2-X_3^2 =L^2\,,
\eeq
which can be parametrized as $X_0=L\cosh \eta_r$ and $\vec{X}=L\,\vec{n}\,\sinh \eta_r$, where $\vec{n}$ is the unit vector.

Working in Milne coordinates $x^\mu=(\tau_r,\eta_r,\theta,\phi)$, we consider the `Hubble' flow
\beq
u_{\tau_r}=-1\,, \qquad u_{\eta_r}=u_\theta=u_\phi=0\,. \label{milflow}
\eeq
 In Minkowski space, this corresponds to
 \beq
 u^\mu = \left(\frac{t}{\tau_r}, \frac{\vec{r}}{\tau_r}\right)\,, \label{hub}
 \eeq
 which has the following properties
\beq
\vartheta=\frac{3}{\tau_r}\,, \qquad \sigma^{\mu\nu}=0\,, \qquad \Omega^{\mu\nu}=0\,.
\eeq
  The flow is similar to Bjorken flow (\ref{bj}) but now the expansion is three-dimensional.
The corresponding energy density in conformal theories is easily obtained as
\beq
\epsilon\propto \frac{1}{\tau_r^{4}}\,. \label{boo}
\eeq
 For more general equations of state $p=w\epsilon$, we have $\epsilon \propto 1/\tau_r^{3(1+w)}$.

\subsubsection{Rotating Hubble flow and the `hybrid' coordinates}
\label{roth}

We now show that the above solution can be generalized to include rotation. Working in the ${\mathbb R}\times H_3$ coordinates $\hat{x}^\mu=(\chi,\eta_r,\theta,\phi)$,   we turn on the $\phi$-component of the velocity for a fluid rotating around the $z$-axis
\beq
\hat{u}_\chi=-\frac{1}{\sqrt{1-\omega^2 \sinh^2\eta_r \sin^2\theta}}\,, \qquad \hat{u}_\phi=\frac{\omega \sinh^2\eta_r \sin^2\theta}{\sqrt{1-\omega^2\sinh^2\eta_r\sin^2\theta}}\,. \label{rot}
\eeq
This still satisfies $\hat{\vartheta}=\hat{\sigma}^{\mu\nu}=0$ but the vorticity tensor no longer vanishes $\hat{\Omega}^{\mu\nu}\neq 0$ (see below).
The corresponding flow velocity in Minkowski space is
\beq
u_t=-\frac{t}{\sqrt{\tau_r^2 -\omega^2 x_\perp^2}} \,, \qquad \vec{u}=\frac{\vec{r}-\omega (\vec{r}\times \vec{e}_z)}{\sqrt{\tau_r^2-\omega^2x_\perp^2}}\,. \label{newflow}
\eeq

The energy density is determined by Eq.\ (\ref{ide}).
The $\mu=\eta_r,\theta$ components are nontrivial and they are simultaneously solved by
\beq
\hat{\epsilon} \propto \frac{1}{(1-\omega^2\sinh^2\eta_r \sin^2\theta)^2}\,,
\eeq
or in Minkowski space,
\beq
\epsilon=\frac{\hat{\epsilon}}{\tau_r^4}\propto \frac{1}{(\tau_r^2-\omega^2 x_\perp^2)^2}=\frac{1}{(t^2-r^2-\omega^2x_\perp^2)^2}\,. \label{newflow2}
\eeq
The fluid is confined  in the region $t^2>r^2+\omega^2x_\perp^2$, meaning that it is squeezed in the $x_\perp$ direction by a factor of $\frac{1}{\sqrt{1+\omega^2}}$. To our knowledge, this rotating solution is new. (See, however, a potentially related \emph{non}relativistic solution recently obtained in \cite{Csorgo:2013ksa}.)
\\

 The flow velocity (\ref{rot}) given in the coordinates $\hat{x}^\mu=(\chi,\eta_r,\theta,\phi)$ has one deficiency: it depends on two variables $\eta_r$ and $\theta$. This will be inconvenient when we consider second-order solutions with rotation. Let us  consider the following change of variables
\beq
\gamma \equiv \sinh \eta_r \sin \theta = \frac{x_\perp}{\tau_r}\,, \qquad \tanh\eta=\tanh \eta_r \cos\theta = \frac{z}{t}\,.
\eeq
In terms of these, the metric becomes
\beq
d\hat{s}^2=\frac{ds^2}{\tau_r^2}=-d\chi^2 + \frac{d\gamma^2}{1+\gamma^2} + (1+\gamma^2) d\eta^2+ \gamma^2 d\phi^2\,. \label{newmet}
\eeq
This is a `hybrid' coordinate system in that we use the three-dimensional `proper time' $\tau_r=\sqrt{t^2-r^2}=e^\chi$ together with the one-dimensional `rapidity' $\eta=\tanh^{-1}\frac{z}{t}$.
The flow velocity in this coordinate system depends only on $\gamma$
\beq
\hat{u}_\chi=-\frac{1}{\sqrt{1-\omega^2\gamma^2}}\,, \qquad \hat{u}_\phi=\frac{\omega \gamma^2}{\sqrt{1-\omega^2\gamma^2}}\,. \label{hyb}
\eeq
The advantage of the hybrid coordinates is that the vorticity tensor has only two nonvanishing components
\beq
\hat{\Omega}^{\gamma \chi} =\omega\gamma^2\hat{\Omega}^{\gamma \phi}= \frac{(1+\gamma^2)\omega^2\gamma}{(1-\omega^2\gamma^2)^{3/2}}\,, \label{vor}
\eeq
modulo the trivial anti-symmetry $\hat{\Omega}^{\mu\nu}=-\hat{\Omega}^{\nu\mu}$. (Remember that $\hat{u}_\mu \hat{\Omega}^{\mu\nu}=0$.) In comparison, $\hat{\Omega}^{\mu\nu}$ has twice as many components in the  ${\mathbb R}\times H_3$ coordinates $\hat{x}^\mu=(\chi,\eta_r,\theta,\phi)$.

\subsection{Conformal soliton flow}
\label{soliton}

Finally, we introduce another spherically expanding flow first discovered by Friess \emph{et al.} in Ref.~\cite{Friess:2006kw} and dubbed `conformal soliton flow' (see also, \cite{Figueras:2009iu}). This solution was rediscovered by Nagy \cite{Nagy:2009eq} and also by us in Ref.~\cite{Hatta:2014gqa} using different methods.

\subsubsection{Einstein static universe}
\label{ein}

There are several ways to describe conformal soliton flow.
 The original derivation in \cite{Friess:2006kw} was based on the mapping of Minkowski space onto the so-called Einstein static universe via the coordinate transformation
\beq
t=\frac{L\sin \xi}{\cos\xi+\cos\sigma}\,, \qquad \vec{r}=\frac{L\sin \sigma }{\cos \xi+\cos \sigma}\vec{n}\,.
\eeq
Equivalently,
\beq
\cot \xi=\frac{L^2+r^2-t^2}{2Lt}\,, \qquad \frac{1}{\cos\xi+\cos\sigma} =\frac{\sqrt{(L^2+(r+t)^2)(L^2+(r-t)^2)}}{2L^2}\,.
\eeq
The metric becomes
\beq
ds^2=\frac{L^2}{\left(\cos \xi+\cos\sigma\right)^2} \left(-d\xi^2+d\sigma^2+\sin^2 \sigma d\Omega^2 \right)\,.
\eeq
 This shows that Minkowski space is conformal to $S^1\times S^3$.
Performing a Weyl transformation, we obtain
\beq
d\hat{s}^2 &=& -d\xi^2+d\sigma^2+\sin^2 \sigma d\Omega^2 \nn
 &=& -d\xi^2 + \frac{dR^2}{1-R^2}+R^2d\Omega^2\,, \label{eins}
\eeq
  where $\sin\sigma \equiv R$ in the second line.
  This is a Robertson-Walker type metric with a constant scale factor and positive spatial curvature.
  It is known as Einstein's static universe which is another solution of the Friedmann equation with a cosmological constant.

The hydrostatic solution static in $\xi$ has the flow velocity $\hat{u}_\xi=-1$, $\hat{u}_\sigma=\hat{u}_\theta=\hat{u}_\phi=0$ which has the properties $\hat{\sigma}^{\mu\nu}=\hat{\vartheta}=\hat{\Omega}^{\mu\nu}=0$.
In  Minkowski space, this corresponds to
\beq
u_t&=&\frac{L}{\cos\xi+\cos\sigma} \frac{\partial \xi}{\partial t}\hat{u}_\xi = -\frac{L^2+r^2+t^2}{\sqrt{(L^2+(r+t)^2)(L^2+(r-t)^2)}}\,, \label{flowid}\\
\vec{u}&=&  \frac{L}{\cos\xi+\cos\sigma}\frac{d\xi}{d\vec{x}}\hat{u}_\xi = \frac{2t\vec{r}}{\sqrt{(L^2+(r+t)^2)(L^2+(r-t)^2)}}\nonumber \,.
\eeq
This is a radially expanding spherically symmetric flow with
the expansion rate
\beq
\vartheta = 3\frac{u^r}{r}\,.
\label{exp}
\eeq
Note that Eq.~(\ref{exp}) is not in contradiction with $\hat{\vartheta}=0$ (the flow is static in the $\hat{x}^\mu$-coordinates) because this quantity does not transform homogeneously under Weyl rescaling.

The continuity equation (\ref{ide}) is trivially solved by $\hat{\epsilon}=const.$ and this  corresponds in Minkowski space to
\beq
\epsilon\propto (\cos\xi+\cos\sigma)^4 \propto\frac{1}{(L^2+(r+t)^2)^2 (L^2+(r-t)^2)^2}\,. \label{eneid}
\eeq
Eqs.\ (\ref{flowid}) and (\ref{eneid}) characterize the conformal soliton flow derived in Ref.~\cite{Friess:2006kw}.
We note that the particular solution with $L=0$ was found slightly earlier in Ref.~\cite{Csorgo:2006ax}.

\subsubsection{$AdS_3\times S^1$}
\label{ads3}

In Ref.~\cite{Hatta:2014gqa} we arrived at the same solution from a different route. Consider the following transformation
\beq
d\hat{s}^2 \equiv \frac{ds^2}{x_\perp^2}
 &=&  \frac{-dt^2 + dz^2 + dx_\perp^2}{x_\perp^2} + d\phi^2 \nn
 &=&  -\cosh^2 \rho \, dT^2 + d\rho^2 + \sinh^2 \rho d\bar{\Theta}^2 + d\phi^2\,.\label{conformal}
\eeq
In the second line we have defined
\beq
\tan T = \frac{L^2 +r^2-t^2}{2Lt}\,, \qquad
\cosh\rho = \frac{1}{2Lx_\perp} \sqrt{(L^2+(r+t)^2)(L^2+(r-t)^2)}\,.
\label{nag}
\eeq
The metric in Eq.~(\ref{conformal}) is that of $AdS_3\times S^1$ where $AdS_3$ is the three-dimensional anti-de Sitter space commonly parametrized as
\beq
X_0^2-X_1^2-X_2^2 + X_3^2 =L^2\,. \label{ads}
\eeq
The relation to the Poincar\'e and global coordinates shown in the first and second line of Eq.~(\ref{conformal}), respectively, is
\beq
&&X_0 = L\frac{t}{x_\perp}=L\cosh \rho \cos T\,,  \qquad
X_1=L\frac{z}{x_\perp}=L\sinh\rho \sin \bar{\Theta}\,, \nonumber \\
&&X_2=\frac{L^2 -r^2 +t^2}{2x_\perp}=L\sinh\rho \cos \bar{\Theta}\,, \quad
X_3=\frac{L^2+r^2-t^2}{2x_\perp}=L\cosh\rho \sin T \,.
\eeq


The hydrostatic solution static in $T$ has the flow velocity
\beq
\hat{u}_T=-\cosh\rho\,, \qquad   \hat{u}^\rho=\hat{u}^{\bar{\Theta}}=\hat{u}^\phi=0\,,
\label{hat}
\eeq
 with the property $\hat{\vartheta}=\hat{\sigma}^{\mu\nu}=\hat{\Omega}^{\mu\nu}=0$.
The corresponding flow velocity in the Minkowski coordinates is obtained by
\beq
u_t &=& -x_\perp \frac{dT}{dt}\hat{u}_T = -\frac{L^2+r^2+t^2}{\sqrt{(L^2+(r+t)^2)(L^2+(r-t)^2)}}\,,  \label{nagy} \\
\vec{u}&=& -x_\perp \frac{dT}{d\vec{x}}\hat{u}_T = \frac{2t\vec{r}}{\sqrt{(L^2+(r+t)^2)(L^2+(r-t)^2)}}\,.
\eeq
 in agreement with (\ref{flowid}).
The energy density $\hat{\epsilon}$ is obtained by solving (\ref{ide}) with
  $\hat{\vartheta}=0$. The result is
 \beq
\hat{\epsilon} \propto \frac{1}{\cosh^4 \rho}\,.
\eeq
The corresponding energy density in Minkowski space $\epsilon=\hat{\epsilon}/x_\perp^4$
is the same as (\ref{eneid}).

\subsubsection{$AdS_2 \times S^2$}
\label{ads2}

Instead of transforming to $AdS_3\times S^1$ as in (\ref{conformal}), let us now write
\beq
ds^2 = -dt^2 + dr^2 + r^2d\Omega^2 = r^2 \left(\frac{-dt^2+dr^2}{r^2} + d\Omega^2 \right)\,.
\eeq
This shows that Minkowski space is also conformal to $AdS_2 \times S^2$.
 As before, we switch to global coordinates
\beq
d\hat{s}^2 = \frac{ds^2}{r^2} = -\cosh^2\tilde{\rho} dT^2 + d\tilde{\rho}^2 + d\theta^2 + \sin^2\theta d\phi^2\,, \label{uuu}
\eeq
 where
\beq
\cosh\tilde{\rho} \equiv \frac{1}{2Lr}\sqrt{(L^2+(r+t)^2)(L^2+(r-t)^2)}\,,
\eeq
 and $T$ is the same as in Eq.~(\ref{nag}).
 It is easy to check that the hydrostatic fluid in this space
is equivalent to the conformal soliton flow in Eqs.~(\ref{flowid}) and (\ref{eneid}) in Minkowski space.

\subsubsection{Rotating conformal soliton flow}
\label{rotn}

As in the case of Hubble flow, it is  possible to rotate  the flow velocity in Eq.~(\ref{flowid}) and find an exact axisymmetric solution. This has been first done by Nagy \cite{Nagy:2009eq} without the use of conformal symmetry techniques. As shown in Ref.~\cite{Hatta:2014gqa}, the rotating solution can be easily implemented in the present framework.

 We work in the $AdS_3\times S^1$ coordinates as in Eq.~(\ref{conformal}). Similarly to Eq.~(\ref{rot}), we try
\beq
\hat{u}_T= \frac{-\cosh^2\rho}{\sqrt{\cosh^2\rho -\omega^2}}\,, \
\qquad
\hat{u}_\phi =\frac{\omega}{\sqrt{\cosh^2\rho -\omega^2}}\,, \label{newu}
\eeq
 which still satisfies $\hat{\vartheta}=\hat{\sigma}^{\mu\nu}=0$ but $\hat{\Omega}^{\mu\nu}\neq 0$.
Differently from the Hubble case as shown in Eq.~(\ref{rot}), now there is a constraint $0\leq \omega \leq 1$.
The corresponding flow velocity in Minkowski space is
\beq
u_t&=&-\frac{L^2+r^2+t^2}{\sqrt{(L^2+(t+r)^2)(L^2+(t-r)^2) -4\omega^2L^2x_\perp^2}}\,, \nn
\vec{u}&=&\frac{2t\vec{r} + 2\omega L (\vec{r}\times \vec{e}_z)}{\sqrt{(L^2+(t+r)^2)(L^2+(t-r)^2)-4\omega^2L^2x_\perp^2}}\,.
\eeq
 With this flow velocity, we can integrate over Eq.~(\ref{ide}) and obtain
\beq
\hat{\epsilon}\propto \frac{1}{(\cosh^2\rho -\omega^2)^2}\,, \label{newhat}
\eeq
or in Minkowski space,
\beq
\epsilon \propto
\frac{1}{\bigl((L^2+(t+r)^2)(L^2+(t-r)^2) -4\omega^2L^2 x_\perp^2\bigr)^2}\,.
\eeq


\section{Exact solutions of second-order conformal hydrodynamics}
\label{seco}

Having described various solutions of ideal hydrodynamics, we now turn to the second-order equations (\ref{a}), (\ref{b}), and (\ref{d}). These are nonlinear, coupled partial differential equations involving 9 unknown variables $(\epsilon,u^\mu,\pi^{\mu\nu})$ (the equations are nonlinear because the various transport coefficients are functions of $\epsilon$, see below). One thus needs some ingenious tricks and assumptions to analytically solve them.

In Ref.~\cite{Gubser:2010ze}, Gubser included only the first term of Eq.~(\ref{d}) (i.e., the Navier-Stokes approximation $\pi^{\mu\nu}\approx -2\eta\sigma^{\mu\nu}$) and obtained an exact analytical solution which generalizes the ideal solution in Section \ref{gubser}. Later, the second-order equations for the gradient expansion \cite{Baier:2007ix} (in which the shear stress tensor is not an independent variable, being completely determined by the gradients of energy density and flow) were studied in Ref.~\cite{Gubser:2010ui}. In general, the inclusion of even a single term in Eq.~(\ref{d}) can make the remaining equations vastly more complex because not only the energy density $\epsilon$ but also the flow velocity $u^\mu$ are modified from the ideal ones. However, Refs.~\cite{Gubser:2010ze,Gubser:2010ui} assumed that $u^\mu$ remains the same as in the ideal solution (\ref{gflow}), as it is completely fixed by the symmetries implemented in terms of the geometry of the associated curved space $dS_3\times {\mathbb R}$, and succeeded in solving the resulting Navier-Stokes equation for $\epsilon$. This approach was further pursued in \cite{Marrochio:2013wla} where the authors included the other terms on the first line of Eq.~(\ref{d}) (i.e., the Israel-Stewart approximation) and found semi-analytical solutions as well as an analytical solution in a certain limit. 

In our previous paper \cite{Hatta:2014gqa}, we have constructed exact second-order solutions which generalize the conformal soliton flow solution in Section \ref{soliton} and take into account, in principle, \emph{all} the terms in Eq.~(\ref{d}). As a matter of fact, many of the terms vanish identically thanks to the flow property which gives $\sigma^{\mu\nu}=0$. Moreover, in Weyl-transformed coordinates we also have $\hat{\vartheta}=0$ so that  Eq.~(\ref{d}) drastically simplifies to
\beq
\hat{\pi}^{\mu\nu} = -\frac{\tau_\pi}{\hat{\epsilon}^{1/4}} \hat{\Delta}^\mu_\alpha \hat{\Delta}^\nu_\beta \hat{D} \hat{\pi}^{\alpha\beta}  + \frac{\lambda_1}{\hat{\epsilon}}\hat{\pi}^{\langle \mu}_{\ \ \lambda}\hat{\pi}^{\nu\rangle\lambda}  \ +\frac{\lambda_2}{\hat{\epsilon}^{1/4}}\hat{\pi}^{\langle \mu}_{\ \ \lambda}\hat{\Omega}^{\nu\rangle\lambda}+ \lambda_3\hat{\epsilon}^{1/2}\hat{\Omega}^{\langle \mu}_{\ \  \lambda}\hat{\Omega}^{\nu\rangle \lambda} \,.
\label{definepi}
\eeq
 In Eq.~(\ref{definepi}), we have redefined the transport coefficients so that they are dimensionless and their $\hat{\epsilon}$-dependence is explicitly factored out. This is because these coefficients are dimensionful in the original Minkowski space and are proportional to $\epsilon$ to some power in a conformal theory (e.g., $\lambda_1 \propto \epsilon^{-1}$). After the equations are Weyl-transformed, this is converted to a power of $\hat{\epsilon}$ (e.g., $\lambda_1 \to \lambda_1/\Lambda^4 \propto \hat{\epsilon}^{-1}$).

With these simplifications, (\ref{definepi}) is finally amenable to analytic approaches and this is what has been done in \cite{Hatta:2014gqa}.
 In this and the next sections, we demonstrate that the technique developed in \cite{Hatta:2014gqa} can be straightforwardly applied to obtain a number of new second-order solutions. We first consider the non-rotating fluids in Milne universe (Section \ref{hubble}) and in Einstein static universe (Section \ref{ein}).

\subsection{Milne universe}
\label{milne}

We work in the coordinates $x^\mu=(\tau_r,\eta_r, \theta,\phi)$ defined in Eq.~(\ref{we}).
Since the flow velocity in Eq.~(\ref{milflow}) satisfies $\sigma^{\mu\nu}=\Omega^{\mu\nu}=0$,
the set of equations (\ref{a})--(\ref{d}) reduce to
\beq
&& \partial_{\tau_r} \epsilon + \frac{4}{\tau_r}\epsilon  =0\,, \label{ae} \\
&& 4\epsilon \nabla_{\tau_r} u^\mu + \Delta^{\mu\alpha}\nabla_\alpha \epsilon +3\Delta^\mu_{\ \nu}\nabla_\alpha \pi^{\alpha\nu}=0\,, \label{be} \\
&&\pi^{\mu\nu} = - \frac{\tau_\pi }{\epsilon^{1/4}}   \left(\Delta^\mu_\alpha \Delta^\nu_\beta \nabla_{\tau_r} \pi^{\alpha\beta} +\frac{4 }{\tau_r}\pi^{\mu\nu} \right) + \frac{\lambda_1}{\epsilon} \pi^{\langle \mu}_{\ \ \lambda}\pi^{\nu\rangle\lambda}\,, \label{ce}
\eeq
 where we have inserted $\vartheta=\frac{3}{\tau_r}$. $\tau_\pi$ and $\lambda_1$ are now redefined to be dimensionless, as we explained above. The general solution of Eq.~(\ref{ae}) is
  \beq
  \epsilon=\frac{f(\eta_r,\theta,\phi)}{\tau_r^4}\,. \label{gsol}
  \eeq
 Inserting Eq.~(\ref{gsol}) into Eq.~(\ref{ce}) and assuming that $\pi^{\mu\nu}$ is diagonal, we can solve the resulting nonlinear coupled differential equation  exactly
\beq
 (\pi^{\eta_r}_{\ \eta_r},\pi^{\theta}_{\ \theta}, \pi^{\phi}_{\ \phi})=
 \frac{1}{\tau_r^4}  \frac{\left(\frac{1}{\tau_r}\right)^{\frac{f^{1/4}}{\tau_\pi}}}{c+\frac{\lambda_1}{f}
 \left(\frac{1}{\tau_r}\right)^{\frac{f^{1/4}}{\tau_\pi}}} \times \begin{cases}
 (-1,-1,2)\,, \\
 (-1,2,-1)\,, \\
 (2,-1,-1)\,, \end{cases} \label{look}
 \eeq
 where $c$ is independent of $\tau_r$.
On the other hand, by working out the connection coefficients, we can rewrite
 Eq.~(\ref{be}) in components
\beq
&&\frac{1}{3}\partial_{\eta_r} \epsilon+\partial_{\eta_r} \pi^{\eta_r}_{\ \eta_r} +\coth\eta_r\left( 2\pi^{\eta_r}_{\  \eta_r} - \pi^{\theta}_{\ \theta} -\pi^{\phi}_{\ \phi} \right)=0\,, \nn
&&\frac{1}{3}\partial_\theta \epsilon+\partial_\theta \pi^{\theta}_{\ \theta} + \cot\theta ( \pi^{\theta}_{\ \theta} - \pi^{\phi}_{\ \phi})=0\,, \nn
&&\frac{1}{3}\partial_\phi \epsilon+\partial_\phi \pi^{\phi}_{\ \phi}=0\,.  \label{component}
\eeq
As long as $\lambda_1\neq 0$, Eqs.~(\ref{look}) and (\ref{component}) are compatible only when $c=0$, in which case the dependence on $\tau_\pi$ drops out. (\ref{component}) can then be easily integrated
\beq
\epsilon=\frac{1}{\tau_r^4}\begin{cases} (\sinh \eta_r\sin\theta)^{\frac{9}{\lambda_1-3}} \,, \\
(\sinh \eta_r)^{\frac{9}{\lambda_1-3}} (\sin \theta)^{-\frac{9}{\lambda_1+6}}\,, \\
(\sinh \eta_r)^{-\frac{18}{\lambda_1+6}}\,, \qquad (\lambda_1\neq 3)\,,
\end{cases} \label{threesol}
\eeq
 for the three cases in Eq.~(\ref{look}). (When $\lambda_1=3$, the first two solutions do not exist and the third solution can be multiplied by any function of $\theta$.) We see that, in contrast to the ideal solution $\epsilon \propto 1/\tau_r^4$, the second-order solutions depend on the rapidity $\sinh\eta_r=\frac{r}{\tau_r}$ as well as the polar angle  $\sin\theta=\frac{x_\perp}{r}$. The latter breaks the spherical symmetry of the ideal solution down to axial symmetry.

 Note that depending on the sign and magnitude of $\lambda_1$,\footnote{We are not aware of any argument that fixes the sign of $\lambda_1$. In a particular model considered in \cite{Molnar:2013lta}, $\lambda_1$ (denoted $-\varphi_7$ in \cite{Molnar:2013lta}) is negative, but this may be model dependent. A positive value of $\lambda_1$ was obtained for ${\mathcal N}=4$ super Yang-Mills in \cite{Baier:2007ix}. However, that paper did not essentially distinguish the $\sigma\sigma$, $\sigma\pi$ and $\pi\pi$ terms. Thus it is not clear whether the result of \cite{Baier:2007ix} corresponds to our $\lambda_1$ or some linear combination of  $\tilde{\eta}_3$, $\tau_{\pi\pi}$, and $\lambda_1$. See, also, Ref.~\cite{York:2008rr}.  \label{foot}}
  the solutions (\ref{threesol})  exhibit singularities at $r=0$ and/or $x_\perp=0$. Near  singular points, some regularization, presumably attributable to higher-order effects, will be needed in practice. Away from the singularities, the solutions are well-behaved and locally satisfy the second-order hydrodynamic equations. Similar comments apply to the other solutions to be presented below.

Incidentally, we can also obtain exact solutions of the Israel-Stewart theory which corresponds to setting $\lambda_1=0$ in Eq.~(\ref{ce}). Substituting Eq.~(\ref{look}) with $\lambda_1=0$ into Eq.~(\ref{component}), we  can determine $f(\eta_r,\theta,\phi)$ and $c=c(\eta_r,\theta,\phi)$. The result is that $f$ is a constant and
\beq
(\pi^{\eta_r}_{\ \eta_r}, \pi^{\theta}_{\ \theta} ,  \pi^{\phi}_{\ \phi}) = \frac{1}{\tau_r^4}\left(\frac{A}{\tau_r}\right)^{\frac{f^{1/4}}{\tau_\pi}}\frac{1}{\sinh^3 \eta_r} \begin{cases} \frac{1}{\sin^3\theta}  (-1,-1,2)\,, \\ \frac{1}{ \sin^\frac{3}{2}\theta} (-1,2,-1)\,,\\ (2,-1,-1)\,,
\end{cases}
\eeq
 where $A$ is an integration constant with the dimension of length.
Therefore, in the Israel-Stewart approximation the energy density $\epsilon=\frac{f}{\tau_r^4}$ is unmodified and the shear-stress tensor decays relative to $\epsilon$ as
\beq
\frac{|\pi^{\mu}_{\ \nu}|}{\epsilon} \propto \left(\frac{1}{\tau_r}\right)^{\frac{f^{1/4}}{\tau_\pi}}=e^{-\frac{f^{1/4}}{\tau_\pi}\ln \tau_r}\,.
\eeq
 Note that the decay is not exponential but power-like in proper time $\tau_r$.
 \\

 In Section \ref{roth}, we have introduced the hybrid coordinates as shown in Eq.~(\ref{newmet}). Although the ideal solution (Hubble flow) can be equally described in the hybrid and the Milne coordinates, the second-order solutions constructed via the two spaces turn out to be different. By repeating the same procedure, we find
 \beq
 \epsilon=\frac{1}{\tau_r^4} \begin{cases} \gamma^{\frac{9}{\lambda_1-3}}\,, \\ (1+\gamma^2)^{\frac{9}{2(\lambda_1-3)}}\,, \\
\bigl(\gamma^2(1+\gamma^2)\bigr)^{-\frac{9}{2(\lambda_1+6)}}\,,
 \end{cases} \label{other}
\eeq
 with $\gamma=\sinh\eta_r \sin\theta= \frac{x_\perp}{\tau_r}$. The first solution of (\ref{other}) agrees with the first one of Eq.~(\ref{threesol}) but the other two solutions are new.

\subsection{Einstein static universe}
\label{adssecond}

Next we include the second-order corrections to the conformal soliton flow in Section \ref{soliton}. In fact, this has already been done in Ref.~\cite{Hatta:2014gqa} by mapping the solution onto $AdS_3\times S^1$. However, our point here is to show that, as already indicated by Eq.~(\ref{other}), starting from the same ideal solution one can obtain different second-order solutions by conformally mapping the solution onto different coordinate systems.

We thus work in the Einstein static universe in Eq.~(\ref{eins}).
In the coordinate system $\hat{x}=(\xi,\sigma,\theta,\phi)$ and with the flow velocity $\hat{u}^\mu=\delta^\mu_\xi$, the hydrodynamic equations read
\beq
&&\partial_\xi\hat{\epsilon}=0\,, \\
&&4\hat{\epsilon}\, \hat{\nabla}_\xi\hat{u}^\mu +\hat{\Delta}^{\mu\nu}\hat{\nabla}_\nu\epsilon+3 \hat{\Delta}^{\mu}_\nu \hat{\nabla}_\alpha \hat{\pi}^{\nu\alpha}= 0\,, \label{kov}\\
&&\hat{\pi}^{\mu\nu}= - \frac{\tau_\pi }{\hat{\epsilon}^{1/4}}  \hat{\Delta}^\mu_\alpha \hat{\Delta}^\nu_\beta \hat{\nabla}_{\xi} \hat{\pi}^{\alpha\beta}+ \frac{\lambda_1}{\hat{\epsilon}} \hat{\pi}^{\langle \mu}_{\ \ \lambda}\hat{\pi}^{\nu\rangle\lambda}\,. \label{unc}
\eeq
 Since $\hat{\epsilon}$ does not depend on $\xi$, neither does $\hat{\pi}^{\mu\nu}$, and this means that the term proportional to $\tau_\pi$ vanishes. Eq.~(\ref{unc}) can then be solved as
 \beq
 (\hat{\pi}^{\sigma}_{\ \sigma},\hat{\pi}^{\theta}_{\ \theta}, \hat{\pi}^{\phi}_{\ \phi})=
 \frac{\hat{\epsilon}}{\lambda_1}\times \begin{cases} (-1,-1,2)\,, \\ (-1,2,-1)\,, \\ (2,-1,-1)\,. \end{cases} \label{lar}
 \eeq
 On the other hand, the nontrivial components of Eq.~(\ref{kov}) are
 \beq
 &&\partial_\sigma \hat{\epsilon}+3\left(\partial_\sigma \hat{\pi}^{\sigma}_{\ \sigma}
 +\cot\sigma (2\hat{\pi}^\sigma_{\ \sigma} -\hat{\pi}^\theta_{\ \theta}-\hat{\pi}^\phi_{\ \phi})\right)=0\,, \nn
 &&\partial_\theta \hat{\epsilon} +3\left(\partial_\theta \hat{\pi}^\theta_{\ \theta}+\cot\theta
 (\hat{\pi}^\theta_{\ \theta} -\hat{\pi}^\phi_{\ \phi})\right)=0\,.
 \eeq
 Inserting Eq.~(\ref{lar}) into the equations above, we can easily integrate over the resulting differential equations
 \beq
 \hat{\epsilon}=\begin{cases}
 \left(\sin\sigma\sin\theta\right)^{\frac{9}{\lambda_1-3}}\,,\\
 \left(\sin\sigma\right)^{\frac{3}{\lambda_1-3}} \left(\sin\theta\right)^{-\frac{9}{\lambda_1+6}}\,,\\
 \left(\sin\sigma\right)^{-\frac{18}{\lambda_1+6}} \,.
 \end{cases}
 \eeq
The corresponding energy density in Minkowski space reads
\beq
\epsilon  \propto \begin{cases} \frac{1}{(L^2+(r+t)^2)^2(L^2+(r-t)^2)^2} \left(\frac{L^2x_\perp^2}{(L^2+(r+t)^2)(L^2+(r-t)^2)}\right)^{\frac{9}{2(\lambda_1-3)}}\,,\\ \frac{1}{(L^2+(r+t)^2)^2(L^2+(r-t)^2)^2} \left(\frac{L^2r^2}{(L^2+(r+t)^2)(L^2+(r-t)^2)}\right)^{\frac{9}{2(\lambda_1-3)}}
\left(\frac{r^2}{x_\perp^2}\right)^{\frac{9}{2(\lambda_1+6)}}\,,\\  \frac{1}{(L^2+(r+t)^2)^2(L^2+(r-t)^2)^2} \left(\frac{L^2 r^2}{(L^2+(r+t)^2)(L^2+(r-t)^2)}\right)^{-\frac{9}{\lambda_1+6}}\,.
\end{cases}
\label{alt}
\eeq
The third solution is spherically symmetric, but when $\lambda_1=3$ we can multiply it by any function of $\sin\theta=x_\perp/r$. We note that exactly the same set of solutions as Eq.~(\ref{alt}) is obtained by working in the $AdS_2\times S^2$ coordinates introduced in Section \ref{ads2}.

The first solution of Eq.~(\ref{alt}) is identical to the first solution of Eq.~(13) in  Ref.~\cite{Hatta:2014gqa} obtained via $AdS_3\times S^1$. However, the other two solutions are new. Thus, the lesson of these analyses is that even if the ideal solution is the same, the second-order solutions constructed via different comformal mappings may in general be different.
\\

A characteristic feature common in the solutions discussed in this section (and also in Ref.~\cite{Hatta:2014gqa}) is that they are nonperturbative in $\lambda_1$. This is due to the behavior
\beq
\pi^{\mu}_{\ \nu} \sim \frac{1}{\lambda_1}\epsilon\,, \label{rey}
\eeq
 which typically  arises as a result of solving `self-consistent' equations of the form $\pi \sim \lambda_1 \pi\pi$.
Eq.\ (\ref{rey}) indicates that $\lambda_1$ essentially plays the role of the Reynolds number $\lambda_1 \sim Re\equiv \epsilon/|\pi^{\mu}_{\ \nu}|$. Since the derivation of the constitutive equation (\ref{d}) from the Boltzmann equation \cite{Denicol:2012cn} is based on the expansion in inverse powers of the Reynolds number, at least in the kinetic theory framework, $\lambda_1$ has to be large for the sake of consistency. 
Indeed, as is clearly seen in (\ref{alt}) for example, the second-order solutions reduce to the ideal one (\ref{eneid}) in the limit $|\lambda_1|\to \infty$ (as we remarked in footnote~\ref{foot}, we do not know \emph{a priori} the sign of $\lambda_1$).

Another striking feature of the solutions in Eq.~(\ref{alt}) is that they are \emph{time-reversible}.
A simple look at the energy-momentum conservation equations (\ref{a}) and (\ref{b}) tells us that time-reversal invariance is broken if $\pi_{\mu\nu}$ is odd under this operation. In fact, when $t\to -t$,  the spatial component of the flow velocity changes as $\vec{u} \to -\vec{u}$, while $\vartheta \to -\vartheta$ and $\sigma_{\mu\nu}\to -\sigma_{\mu\nu}$. Thus, in the Navier-Stokes approximation $\pi_{\mu\nu} \sim -2\eta \sigma_{\mu\nu}$, one can clearly see that time-reversal invariance is broken, and this should be associated with the production of entropy.
However, our solutions are dual to a static fluid in the Weyl-transformed space $\hat{x}^\mu$ in which the metric $\hat{g}^{\mu\nu}$ does not depend on `time' $\hat{x}^0$, either. As a result, all the terms that potentially break time-reversal invariance  vanish  $\hat{\sigma}^{\mu\nu}=\hat{\vartheta}=\hat{D}\hat{\pi}^{\mu\nu}=0$. It then follows that $\pi^{\mu\nu}$ must be even under time-reversal, and this is manifest in Eq.~(\ref{alt}).

An immediate consequence of time-reversibility is that entropy is not produced in these solutions even though $\pi^{\mu\nu}\neq 0$. This is in contradiction to the pragmatic definition of the `nonequilibrium entropy' commonly employed in the literature  (e.g., Ref.~\cite{DeGroot:1980dk})
\beq
s_{noneq} \equiv s-\frac{\tau_\pi}{4\eta T}  \pi_{\mu\nu}\pi^{\mu\nu}\,, \label{ent}
\eeq
where $T$ here is the temperature and $s=\frac{\epsilon+p}{T}$ is the equilibrium entropy. Eq.~(\ref{ent}) implies that whenever $\pi^{\mu\nu}$ is nonvanishing, there is an associated entropy production
\beq
\partial_\mu (s_{noneq}u^\mu) = \frac{1}{2\eta T}\pi_{\mu\nu}\pi^{\mu\nu}+\cdots\,.
\eeq
However, our findings suggest a potential flaw in this argument.  For other definitions of $s_{noneq}$, see for instance   Refs.~\cite{Loganayagam:2008is,Romatschke:2009kr,Bhattacharyya:2012nq}.

\section{Second-order solutions with rotation}
\label{rotation}

In the previous section, we constructed solutions of second-order hydrodynamic equations for irrotational flows, namely, flows with vanishing vorticity $\Omega^{\mu\nu}=0$. In Ref.~\cite{Hatta:2014gqa}, we have for the first time found  a second-order solution with $\Omega^{\mu\nu}\neq 0$ by generalizing Nagy's rotating solution in Section \ref{rotn}. Here we construct another rotating second-order solution starting from the rotating Hubble flow derived in Section \ref{roth}. The calculations turn out to be very similar to the former case, therefore the readers may find the present section as a helpful guide to follow the exposition in Ref.~\cite{Hatta:2014gqa}.

We work in the hybrid coordinates in Eq.~(\ref{newmet}) $\hat{x}^\mu=(\chi,\gamma,\eta,\phi)$ in which the rotating flow velocity takes the form as in Eq.~(\ref{hyb}). The vorticity tensor in Eq.~(\ref{vor}) induces a new term
\beq
\hat{\Omega}^{\langle\mu}_{\ \ \lambda}\hat{\Omega}^{\nu\rangle \lambda}=\frac{ (\hat{\Omega}^{\gamma\phi})^2}{3}\begin{pmatrix} \frac{\omega^2\gamma^4}{1+\gamma^2} & 0 & 0 & \frac{\omega \gamma^2}{1+\gamma^2} \\  0 & \gamma^2(1-\omega^2\gamma^2) & 0 & 0 \\
0 & 0 & -\frac{2\gamma^2 (1-\omega^2\gamma^2)}{(1+\gamma^2)^2} & 0  \\
 \frac{\omega \gamma^2}{1+\gamma^2} & 0 & 0 & \frac{1}{1+\gamma^2}
\end{pmatrix}\,, \label{ooomega}
\eeq
on the right-hand-side of the constitutive equation (\ref{definepi}).

In order to solve Eq.~(\ref{definepi}), let us temporarily assume that $\tau_\pi=\lambda_2=0$ (we shall relax this assumption shortly). Unlike the non-rotating cases, $\hat{\pi}^{\mu\nu}$ cannot be diagonal due to the condition $\hat{u}_\mu \hat{\pi}^{\mu\nu}=0$, but one can make the simplest Ansatz, which takes the form
\beq
\hat{\pi}^{\mu\nu}= \begin{pmatrix} \omega^2\gamma^4 \hat{\pi}^{\phi\phi} & 0 & 0 & \omega\gamma^2\hat{\pi}^{\phi\phi} \\ 0 & \hat{\pi}^{\gamma\gamma} & 0 & 0 \\
0 & 0 & \hat{\pi}^{\eta\eta} & 0 \\
\omega\gamma^2 \hat{\pi}^{\phi\phi} & 0 & 0 & \hat{\pi}^{\phi\phi}
\end{pmatrix}\,. \label{pipi}
\eeq
Substituting Eqs.~(\ref{ooomega}) and (\ref{pipi}) into Eq.~(\ref{definepi}), we find a set of equations
 \beq
 X&=&\frac{\lambda_1}{\hat{\epsilon}} \left(X^2-\frac{X^2+Y^2+Z^2}{3}\right)+ \frac{\lambda_3\sqrt{\hat{\epsilon}}}{3} \frac{\omega^2 (1+\gamma^2)}{(1-\omega^2\gamma^2)^2}\,, \nn
 Y&=& \frac{\lambda_1}{\hat{\epsilon}} \left(Y^2-\frac{X^2+Y^2+Z^2}{3}\right)- \frac{2\lambda_3\sqrt{\hat{\epsilon}}}{3} \frac{\omega^2 (1+\gamma^2)}{(1-\omega^2\gamma^2)^2}\,, \nn
 Z&=& \frac{\lambda_1}{\hat{\epsilon}} \left(Z^2-\frac{X^2+Y^2+Z^2}{3}\right)+ \frac{\lambda_3\sqrt{\hat{\epsilon}}}{3} \frac{\omega^2 (1+\gamma^2)}{(1-\omega^2\gamma^2)^2}\,,
 \label{four}
  \eeq
 where
 \beq
 X=\hat{\pi}^{\gamma}_{\ \gamma}\,, \qquad Y=\hat{\pi}^{\eta}_{\ \eta}\,, \qquad
 Z=(1-\omega^2\gamma^2)\hat{\pi}^{\phi}_{\ \phi}\,.
 \eeq
 Eq.\ (\ref{four}) admits four solutions
 \beq
 && X=Z=-\frac{Y}{2} = \frac{\hat{\epsilon}}{2\lambda_1}\left(-1\pm \sqrt{1+\frac{4f}{3}}\right)\,,
 \label{same} \\
 && (X,Y,Z)=\frac{\hat{\epsilon}}{\lambda_1}\left(\frac{1}{2}\left(1\pm \sqrt{9-4f}\right), -1, \frac{1}{2}(1\mp \sqrt{9-4f})\right)\,, \label{dif}
 \eeq
  where we defined
  \beq
  f\equiv \frac{\lambda_1\lambda_3 \omega^2(1+\gamma^2)}{\sqrt{\hat{\epsilon}}(1-\omega^2\gamma^2)^2}\,. \label{fw}
  \eeq
 It turns out that the first two solutions in Eq.~(\ref{same}) satisfy $\hat{\Delta}^\mu_\alpha \hat{\Delta}^\nu_\beta \hat{D} \hat{\pi}^{\alpha\beta}=\hat{\pi}^{\langle \mu}_{\ \ \lambda}\hat{\Omega}^{\nu\rangle\lambda}=0$, i.e., they are solutions even when $\tau_\pi, \, \lambda_2\neq 0$. We thus consider only the solutions in Eq.~(\ref{same}) in the following.\footnote{In fact, if $\tau_\pi$ and $\lambda_2$ are related as $\tau_\pi=-2\lambda_2$,  the last two solutions (\ref{dif}) are also acceptable solutions because the two terms cancel exactly $-\tau_\pi \hat{\Delta}^\mu_\alpha \hat{\Delta}^\nu_\beta \hat{D} \hat{\pi}^{\alpha\beta}+\lambda_2\hat{\pi}^{\langle \mu}_{\ \ \lambda}\hat{\Omega}^{\nu\rangle\lambda}=0$. However, we have no reason to believe that this relation generally holds. Note that it differs from the kinetic theory prediction $\tau_\pi=2\lambda_2$  by a minus sign.}

We now look at the equation for $\hat{\epsilon}$
\beq
4\hat{\epsilon}\, \hat{D}\hat{u}^\mu +\hat{\Delta}^{\mu\nu}\hat{\nabla}_\nu\epsilon+3 \hat{\Delta}^{\mu}_\nu \hat{\nabla}_\alpha \hat{\pi}^{\nu\alpha}&=& 0\,.
\eeq
The $\mu=\gamma$ component is nontrivial and reads
\beq
(1+\gamma^2) \partial_\gamma\hat{\epsilon} -\frac{4\omega^2 \gamma (1+\gamma^2)}{1-\omega^2\gamma^2} \hat{\epsilon} + 3\left(\partial_\gamma \hat{\pi}^{\gamma\gamma} + \frac{\hat{\pi}^{\gamma\gamma}}{\gamma (1+\gamma^2)} -\gamma (1+\gamma^2)(\hat{\pi}^{\eta\eta}+\hat{\pi}^{\phi\phi}) \right)\,.
\eeq
After inserting Eq.~(\ref{same}) into the above equation, we find
\beq
\partial_\gamma \hat{\epsilon} -\frac{4\gamma\omega^2}{1-\omega^2\gamma^2}\hat{\epsilon}
+3\left(\partial_\gamma X -\frac{4\gamma\omega^2}{1-\omega^2\gamma^2}X \right)+\frac{9\gamma(1+\omega^2)}{(1-\omega^2\gamma^2)(1+\gamma^2)}X=0\,. \label{diffic}
\eeq
Due to the $\hat{\epsilon}$ dependence in $X$ through Eq.~(\ref{fw}), Eq.~(\ref{diffic}) is a complicated nonlinear differential equation for $\hat{\epsilon}$ which is difficult to solve.
We thus employ an Ansatz
\beq
\hat{\epsilon}= \frac{A^2(\gamma)(1+\gamma^2)^2}{(1-\omega^2\gamma^2)^4}\,, \label{ansa}
\eeq
 with which we can write $X=b(\gamma)\hat{\epsilon}$ (cf., (\ref{same})) where $b$ is the root of
 \beq
 |A(\gamma)|=\frac{\lambda_3\omega^2}{3b(\gamma)(\lambda_1 b(\gamma)+1)}\,. \label{AA}
 \eeq
  Using Eq.~(\ref{same}), one can check that the  ratio
 $\lambda_3/b(\lambda_1 b+1)$ is positive.
 However, {\it a priori} we do not know the sign of $\lambda_3$.

Let us first assume that $A$ is a constant. In this case, Eq.~(\ref{diffic}) reduces to $(1+\omega^2)(4+21b)=0$, meaning that $b=-\frac{4}{21}$ and
\beq
|A|=\frac{7\lambda_3\omega^2}{4\left(\frac{4}{21}\lambda_1-1 \right)}\,.
\eeq
  Since the right-hand-side must be positive, this solution exists only when $\lambda_1>\frac{21}{4}$ if $\lambda_3$ is positive, and $\lambda_1<\frac{21}{4}$ if $\lambda_3$ is negative. If we take the limits $\omega\to 0$ and $\lambda_1\to \frac{21}{4}$  such that $A$ remains finite,  Eq.~(\ref{ansa}) reduces to the second solution in Eq.~(\ref{other}).

When $A$ is not a constant, we find a differential equation for $b(\gamma)$
\begin{equation}
\frac{db(\gamma)}{d\gamma}\left(\frac{9\lambda_1 b^2+4\lambda_1 b + 3b+2}{b (\lambda_1 b+1)(4+21b)}\right)=
\frac{\gamma(1+\omega^2)}{(1-\omega^2\gamma^2)(1+\gamma^2)}\,.
\end{equation}
This can be integrated by separation of variables. The result is
\begin{equation}
b(\lambda_1 b+1) \left|1+\frac{21}{4}b\right|^{e_1(\lambda_1)}|1+\lambda_1 b|^{e_2(\lambda_2)} = \lambda_3 \omega^2C\frac{1+\gamma^2}{1-\omega^2\gamma^2}\,. \label{can}
\end{equation}
where $e_1(\lambda_1)\equiv \frac{105-32\lambda_1}{7(4\lambda_1-21)}$,  $e_2(\lambda_1)\equiv \frac{9}{4\lambda_1-21}$ and $C>0$ is the integration constant.\footnote{In order for $\epsilon$ in Eq.~(\ref{thus}) to have a finite limit as $\lambda_3 \to 0$ or $\omega\to 0$, the integration constant must be proportional to $\lambda_3\omega^2$ which is explicitly factored out in Eq.~(\ref{can}). We also used the fact that $\lambda_3/b(\lambda_1 b+1)$ is positive, as already remarked.}
We thus find the energy density
\begin{equation}
 \epsilon=\frac{\hat{\epsilon}}{\tau_r^4} = \frac{1}{9C^2(t^2-r^2-\omega^2 x_\perp^2)^2} \left| 1+\frac{21}{4}b(\gamma)\right|^{2e_1(\lambda_1)} | 1+\lambda_1 b(\gamma)|^{2e_2(\lambda_1)} \,, \label{thus}
\end{equation}
 where $b$ is the solution of Eq.~(\ref{can}). It is useful and convenient to define the quantity
\beq
R(\gamma)\equiv \left| 1+\frac{21}{4}b(\gamma)\right|^{2e_1(\lambda_1)} |1+\lambda_1 b(\gamma)|^{2e_2(\lambda_1)}\,.
\label{deviationR}
\eeq
The deviation of $R(\gamma)$ from unity is a local measure of the strength of the second-order effects at the corresponding spacetime point $\gamma^2= \frac{x^2_\perp}{t^2-z^2-x_\perp^2}$, since $R(\gamma) \to 1$ in the $\lambda_1\to \infty$ limit.

For a given value of $\gamma$, one can solve Eq.~(\ref{can}) for $b$ numerically and the energy density at that point $\epsilon(t,x_\perp,z)=\epsilon(\gamma)$ is determined from  Eq.~(\ref{thus}). Actually, depending on the sign of $\lambda_1$ and $\lambda_3$, Eq.~(\ref{can}) admits multiple solutions. Not all of them are physically acceptable because we must ensure that $b = \hat{\pi}^{\gamma}_{\ \gamma}/\hat{\epsilon}$ should be bounded as $\gamma$ is varied between 0 and $1/\omega$. Also, $R$ should not be much larger or much smaller than unity at least in some region of $\gamma$, otherwise the second-order effects are too large to be reliable. This puts constraints on the relative size of the parameters involved.

\begin{figure*}[ht]
\includegraphics[width=8cm]{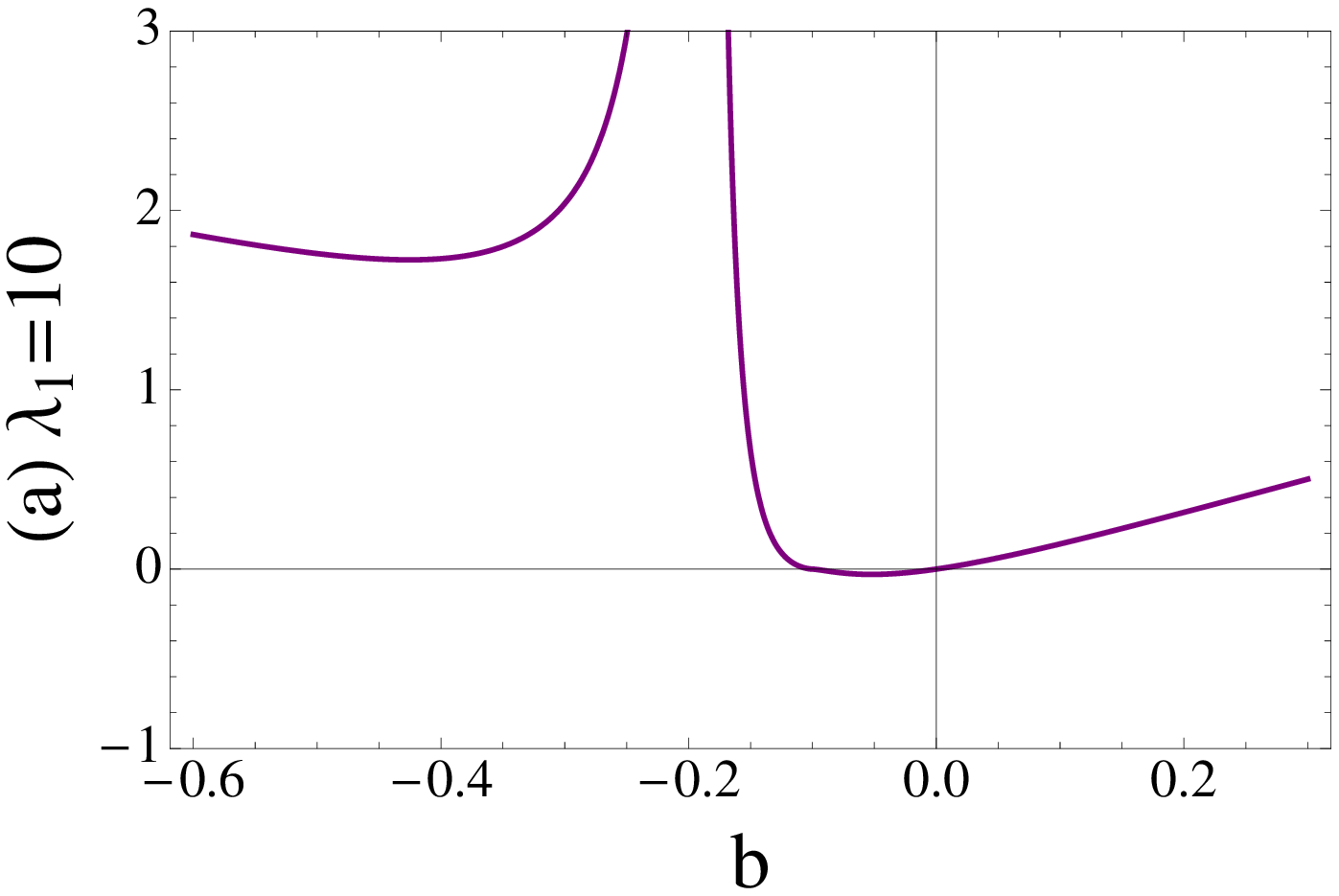}\quad \includegraphics[width=8cm]{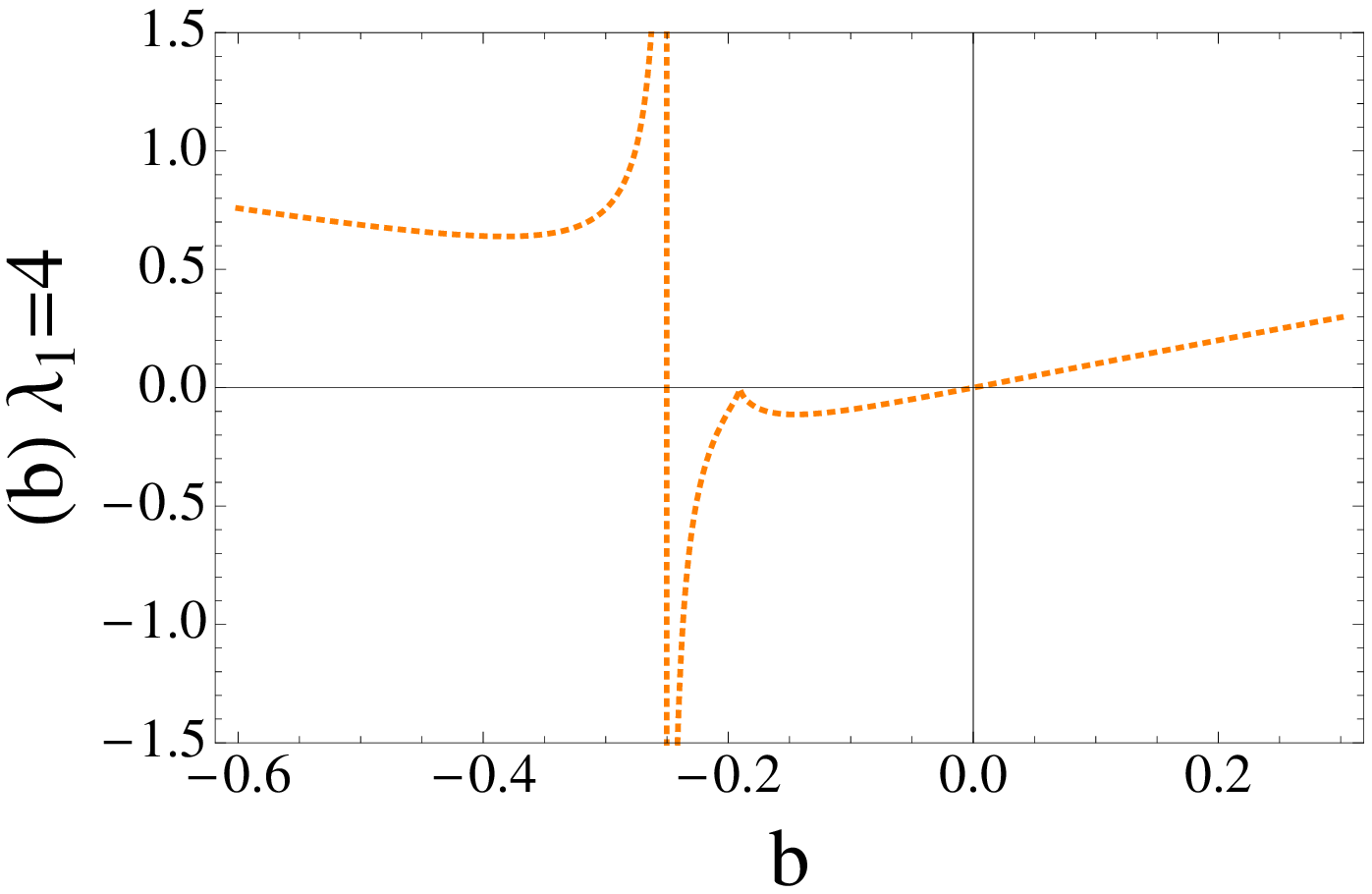} \\
\includegraphics[width=8cm]{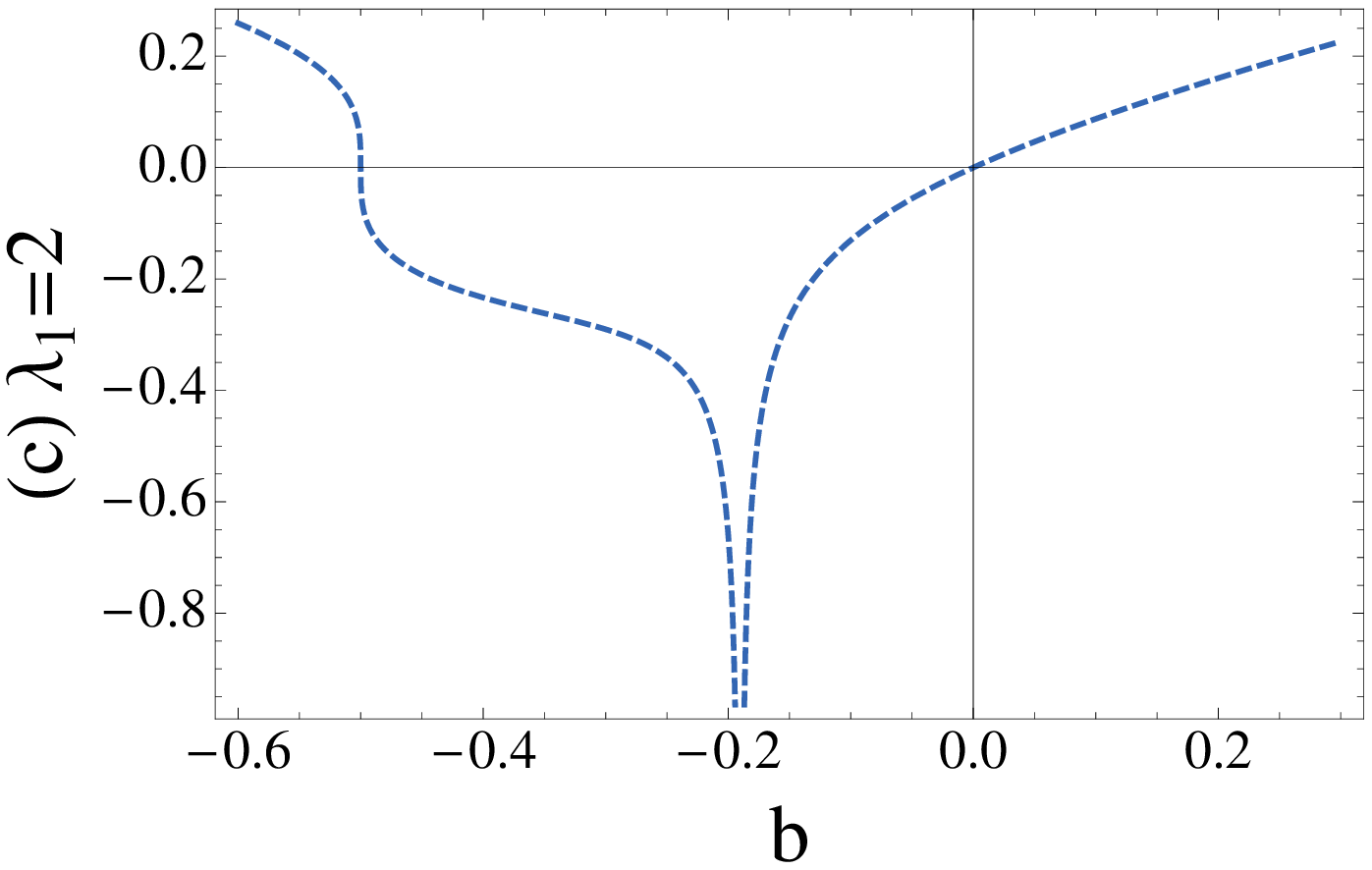}\quad \includegraphics[width=8cm]{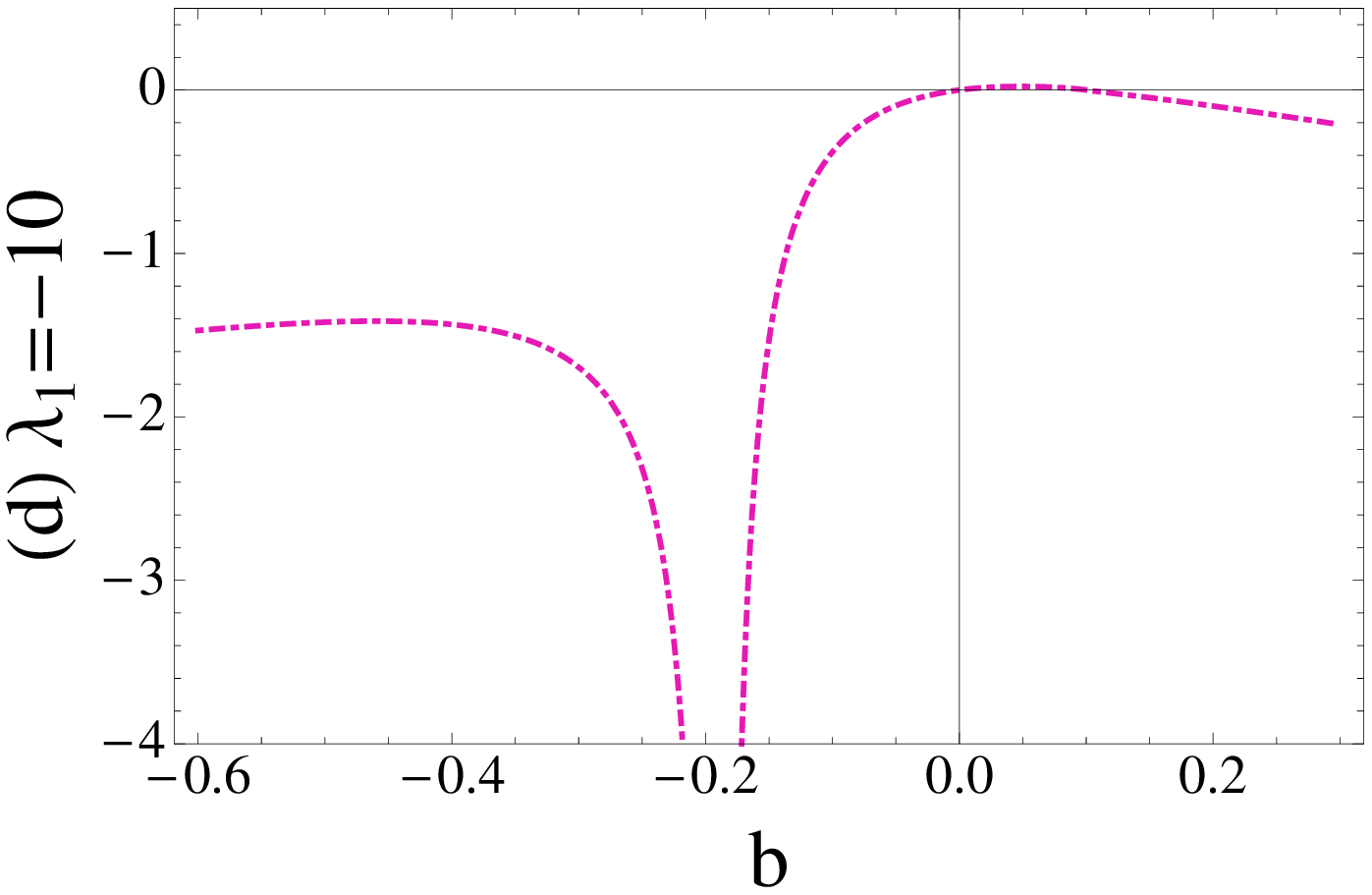}
\caption{\label{fig1} (Color online) Plots of the left-hand-side of (\ref{can}) with different values of $\lambda_1$.}
\end{figure*}

\begin{figure*}[ht]
\includegraphics[width=12cm]{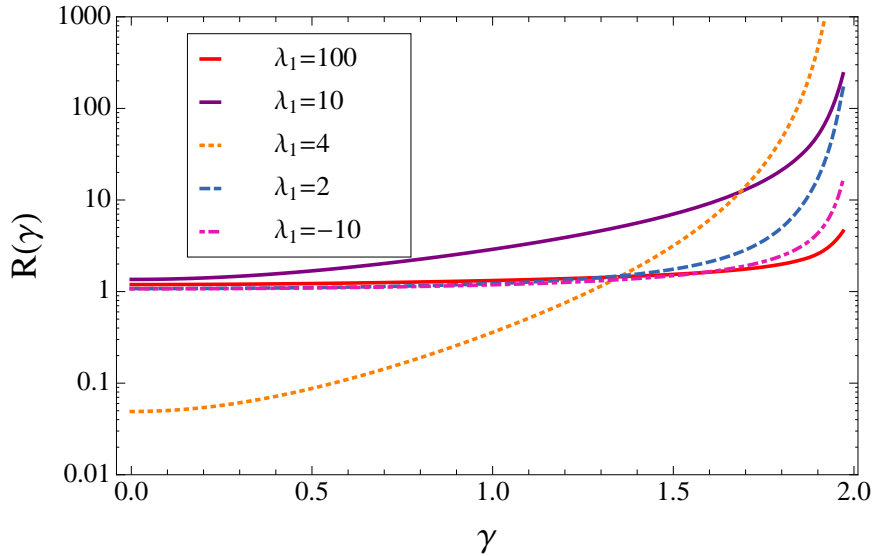}
\caption{\label{fig2} (Color online) Deviation from the ideal solution defined in (\ref{deviationR}) for $\omega=0.5$ (so that $2>\gamma\ge 0$) and $\lambda_3\omega^2C=\pm 0.01$, depending on the sign of $\lambda_3$. Solid curves are for positive values of $\lambda_3$, while the dashed and dotted curves are for negative values of $\lambda_3$. The drawing and color of the curves in this figure are correlated with the curves shown in Fig.~\ref{fig1}.}
\end{figure*}

 Without knowing the sign of $\lambda_1$ and $\lambda_3$ (see footnote \ref{foot}),  we find it necessary to consider three different regions of $\lambda_1$ separately:

  \begin{itemize}
  \item $\lambda_1>\frac{21}{4}$ \\
   The left-hand-side of Eq.~(\ref{can}) is plotted in Fig.~\ref{fig1}(a) for $\lambda_1=10$.  Clearly, $\lambda_3$ has to be positive and the roots of Eq.~(\ref{can}) can be found in the regions $b>0$ and $b<-1/\lambda_1$. The solution at $b>0$ should be discarded because $b\to +\infty $ as $\gamma\to 1/\omega$. We find an acceptable solution $b(\gamma)$ which varies monotonously in the region $-1/\lambda_1 > b > -4/21$. The corresponding $R(\gamma)$, namely Eq.~(\ref{deviationR}), is plotted in Fig.~\ref{fig2} for $\lambda_1=10$ and $\lambda_1=100$. If $C$ is not too small, there is a second branch of solutions in which $b$ starts from the local minimum at $b<-4/21$ ($b\approx -0.4$ in Fig.~\ref{fig1}(a)) and asymptotically approaches $b\to -4/21$ as $\gamma\to 1/\omega$. However, for this solution $R$ becomes larger than unity by several orders of magnitude, so we discard it as an unphysical solution.

    \item $\frac{21}{4}>\lambda_1 >\frac{105}{32}$  \\
   Fig.~\ref{fig1}(b) shows the left-hand-side of (\ref{can}) for $\lambda_1=4$. If $\lambda_3$ is positive and $C$ is not too small, there is a solution that starts from the local minimum at $b<-1/\lambda_1$ and asymptotically approaches $-1/\lambda_1$ from below as $\gamma\to 1/\omega$. On the other hand, if $\lambda_3$ is negative there is another branch of solutions in which $b$ varies in the interval $-4/21>b>-1/\lambda_1$. However, for these solutions $R(\gamma)$ tends to become very large towards the boundary of the fluid $\gamma\to 1/\omega$. The result for $\lambda_1=4$, $\lambda_3<0$ is plotted in Fig.~\ref{fig2}.

    \item    $\frac{105}{32}>\lambda_1$ \\
  Fig.~\ref{fig1}(c) and (d) show the left-hand-side of (\ref{can}) for $\lambda_1=2$ and $\lambda_1=-10$, respectively.
    For $\lambda_3<0$, we always have an acceptable solution that starts from $b\approx 0$ and asymptotically approaches $b=-4/21$. This is plotted in Fig.~\ref{fig2} for $\lambda_1=2$ and $\lambda_1=-10$. Moreover, there could be other branches of solutions. However, as discussed above, they have too large values of $R$ to be acceptable. 


\end{itemize}

In all the acceptable solutions, we observe the clear tendency that $R$ becomes large towards the boundary of the rotating fluid $\gamma\to 1/\omega$. Actually the ideal solution itself becomes very large near the boundary, and the vorticity effect further amplifies the growth there irrespective of the sign of $\lambda_3$.

\section{Unorthodox Bjorken flow}
\label{shear}

All the solutions discussed so far have vanishing shear-tensor ($\sigma^{\mu\nu}=0$), which is rather special.
When $\sigma^{\mu\nu}$ is nonzero, the constitutive equation in its most general form as shown in Eq.~(\ref{d}) is obviously much more difficult to solve even with the help of conformal symmetry. Nevertheless, here we revisit the boost-invariant setup, which was briefly reviewed in Section \ref{bjo}, as the simplest example with $\sigma^{\mu\nu}\neq 0$, and demonstrate that one can still find exact scaling solutions.

As a warm-up, let us consider the effect of shear viscosity on Bjorken flow in a conformal theory. The continuity equation (\ref{a}) takes the form
 \beq
 \partial_\tau \epsilon + \frac{4}{3\tau}\epsilon + \frac{1}{3\tau}(2\pi^\eta_{\ \eta}-\pi^x_{\ x} -\pi^y_{\ y})=0\,,  \label{eneb}
\eeq
 where we have used Eq.~(\ref{compo}). In the Navier-Stokes approximation, $\pi^{\mu\nu}=-2\eta \sigma^{\mu\nu}$ with $\sigma^{\mu\nu}$  given by Eq.~(\ref{compo}). Substituting this into Eq.~(\ref{eneb}) and noticing that $\eta\propto \epsilon^{3/4}$, we arrive at
\beq
\partial_\tau \epsilon +\frac{4}{3\tau}\epsilon -\frac{4\eta}{3\tau^2} \epsilon^{3/4}=0\,, \label{varb}
\eeq
 where we  redefined $\eta>0$ to be dimensionless. A common practice to solve (\ref{varb}) is via a perturbative expansion. Recalling that the ideal solution (\ref{bjorken}) scales as $\epsilon \propto \tau^{-4/3}$, one finds the asymptotic solution at large-$\tau$
 \beq
 \epsilon = \frac{c}{\tau^{4/3}} - \frac{2\eta c^{3/4}}{\tau^2} + {\mathcal O}(\tau^{-8/3})\,, \label{accept}
 \eeq
 where $c>0$ is arbitrary. However, it is not clear a priori what is the radius of convergence of such a series \cite{Denicol:2011fa}. In fact, a recent study of the large order behavior of the hydrodynamic gradient expansion at strong coupling has found the radius of convergence to be zero \cite{Heller:2013fn}, which is characteristic of an asymptotic series. In addition, for non-linear differential equations in general, there could be solutions which can never be reached by doing perturbative expansions.

 Are there other solutions to Eq.~(\ref{varb})? Let us try the Ansatz
 \beq
 \epsilon = \frac{C}{\tau^\alpha}\,. \label{ansatz}
 \eeq
Substituting this into (\ref{varb}), we find $\alpha=4$ together with the relation $C^{1/4}=-\frac{\eta}{2}$, but the latter is not physically acceptable because the fluid temperature must be positive $T\sim\frac{C^{1/4}}{\tau}>0$. If it were not for this sign mismatch, however, the rather naive choice (\ref{ansatz}) would have become an exact solution. This is due to conformal symmetry. Since the only dimensionful parameter in Eq.~(\ref{varb}) is $\tau$, $\epsilon\propto \tau^{-4}$ is a natural guess based on the ground of dimensional analysis\footnote{In the Bjorken solution $\epsilon \propto \tau^{-4/3}$ of ideal hydrodynamics, an additional dimensionful parameter is provided by the initial condition.} (though the flow expansion looks like three-dimensional rather than one-dimensional, cf., (\ref{boo})).

We now include the second-order terms. The conservation equation (\ref{eneb}) is unchanged
 but now $\pi^{\mu\nu}$ must be determined by solving Eq.~(\ref{d}) with
  $\Omega^{\mu\nu}=0$. Rescaling the transport coefficients as we have done before, we arrive at the following equation
 \beq
\pi^{\mu\nu} &=&-2\eta \epsilon^{3/4}\sigma^{\mu\nu} - \frac{\tau_\pi}{\epsilon^{1/4}} \left(\Delta^\mu_\alpha \Delta^\nu_\beta  D\pi^{\alpha\beta}+\frac{4}{3}\vartheta \pi^{\mu\nu}\right) \nn
&& + \frac{\lambda_1}{\epsilon} \pi^{\langle \mu}_{\ \ \lambda}\pi^{\nu\rangle\lambda}    +\tau_\sigma \epsilon^{1/2} \left(\Delta^\mu_\alpha \Delta^\nu_\beta  D\sigma^{\alpha\beta}+\frac{1}{3} \sigma^{\mu\nu}\vartheta\right) \nn
&& -\tilde{\eta}_3 \epsilon^{1/2}\sigma^{\langle \mu}_{\ \  \lambda}\sigma^{\nu\rangle \lambda}-\frac{\tau_{\pi\pi}}{\epsilon^{1/4}} \sigma^{\langle \mu}_{\ \  \lambda}\pi^{\nu\rangle \lambda}\,. \label{full}
\eeq
With the properties as shown in Eq.~(\ref{compo}) at hand, we can write down explicitly all the non-trivial components of Eq.~(\ref{full}) as follows
\beq
 \pi^{\eta}_{\ \eta} &=&-\frac{4\eta \epsilon^{3/4}}{3\tau} -\frac{\tau_\pi}{\epsilon^{1/4}}\left(\partial_\tau \pi^\eta_{\ \eta}+\frac{4}{3\tau}\pi^\eta_{\ \eta}\right) + \frac{\lambda_1}{3\epsilon} \bigl( 2(\pi^\eta_{\ \eta})^2 -(\pi^{x}_{\ x})^2-(\pi^y_{\ y})^2\bigr)\,, \nn
&&\qquad -2\frac{2\tau_\sigma+\tilde{\eta}_3}{9\tau^2}\epsilon^{1/2} -\frac{\tau_{\pi\pi}}{3\tau\epsilon^{1/4}}\pi^{\eta}_{\ \eta}\nn
\pi^{x}_{\ x} &=&\frac{2\eta \epsilon^{3/4}}{3\tau} -\frac{\tau_\pi}{\epsilon^{1/4}}\left(\partial_\tau \pi^x_{\ x}+\frac{4}{3\tau}\pi^x_{\ x}\right) + \frac{\lambda_1}{3\epsilon} \bigl( 2(\pi^x_{\ x})^2 -(\pi^{\eta}_{\ \eta})^2-(\pi^y_{\ y})^2\bigr)\,, \nn
&& \qquad  +\frac{2\tau_\sigma+\tilde{\eta}_3}{9\tau^2}\epsilon^{1/2} -\frac{\tau_{\pi\pi}}{3\tau\epsilon^{1/4}}\pi^{y}_{\ y} \nn
\pi^{y}_{\ y} &=&\frac{2\eta \epsilon^{3/4}}{3\tau} -\frac{\tau_\pi}{\epsilon^{1/4}}\left(\partial_\tau \pi^y_{\ y}+\frac{4}{3\tau}\pi^y_{\ y}\right) + \frac{\lambda_1}{3\epsilon} \bigl( 2(\pi^y_{\ y})^2 -(\pi^{\eta}_{\ \eta})^2-(\pi^x_{\ x})^2\bigr) \nn
&&  \qquad +\frac{2\tau_\sigma+\tilde{\eta}_3}{9\tau^2}\epsilon^{1/2} -\frac{\tau_{\pi\pi}}{3\tau\epsilon^{1/4}}\pi^{x}_{\ x}
 \,. \label{vbv}
 \eeq
Eq.~(\ref{vbv}) may be solved perturbatively as in (\ref{accept}). Here instead, we look for nonperturbative \emph{exact} solutions. For this purpose, we again try the Ansatz as shown in Eq.~(\ref{ansatz}) with $\alpha=4$. Eq.~(\ref{eneb}) then requires that
 \beq
 2\pi^\eta_{\ \eta}-\pi^x_{\ x} -\pi^y_{\ y}=8\epsilon\,. \label{take}
 \eeq
 Since $\pi^\eta_{\ \eta}+\pi^x_{\ x}+\pi^y_{\ y}=0$, (\ref{take}) gives 
 \beq
 \pi^{\eta}_{\ \eta}= \frac{8}{3}\epsilon=\frac{8C}{3\tau^4}\,.
 \eeq
 Substituting this into Eq.~(\ref{vbv}), after some algebra, we find the following two sets of exact solutions
\beq
 && \pi^x_{\ x}=\pi^y_{\ y}=-\frac{4}{3}\epsilon\,,
\nn
&& C^{1/4}= \frac{3\eta-16\tau_\pi +2\tau_{\pi\pi} \pm \sqrt{ (3\eta-16\tau_\pi +2\tau_{\pi\pi})^2+4(4\lambda_1-3)(2\tau_\sigma+\tilde{\eta}_3)}}{4(4\lambda_1-3)}\,, \label{C4}
\eeq
  and
\beq
&& \left(\frac{\pi^x_{\ x}}{\epsilon}, \frac{\pi^y_{\ y}}{\epsilon}\right)= -\frac{4}{3} \pm \sqrt{16-\frac{1}{\lambda_1}
\left(\frac{2(3\eta+4\tau_{\pi\pi})}{3C^{1/4}} +\frac{2\tau_\sigma+\eta_3}{3C^{1/2}}\right)}\,, \label{squ} \nn
&&C^{1/4}=\frac{8\tau_\pi+\tau_{\pi\pi}}{8\lambda_1+3}\,. \label{C42}
\eeq
These solutions make sense as long as $C^{1/4}$ is positive (for the second solution we also need to require that the quantity in the square-root is positive). This was not the case in the Navier-Stokes approximation, but we now see that there are regions in the parameter space where this is possible. Note that in these solutions, there is no integration constant. The overall normalization is completely fixed due to the nonlinearity of the equation. While this may be an unattractive feature from a phenomenological viewpoint,  the solutions are still remarkable as they explicitly depend on \emph{six} different transport coefficients!

As already remarked in Section \ref{secondorder}, in the literature, one often treats $\pi^{\mu\nu}$ and $-2\eta \sigma^{\mu\nu}$ interchangeably in the equations of second-order hydrodynamics after which there is no distinction between $D\pi$ and $D\sigma$ terms, or among $\pi\pi$, $\pi\sigma$ and $\sigma\sigma$ terms. Therefore, only one term from the `degenerate' set of terms is kept. However, we see that the respective transport coefficients $\tau_\pi$ and $\tau_\sigma$, or $\lambda_1$,  $\tau_{\pi\pi}$ and $\tilde{\eta}_3$ enter differently in  Eqs~(\ref{C4}) and (\ref{C42}), with the coefficients of the gradient terms ($\tau_\sigma,\tilde{\eta}_3,\tau_{\pi\pi}$) playing relatively minor roles.  Accordingly, in the above solutions the relation $\pi^{\mu\nu}\approx -2\eta \sigma^{\mu\nu}$ is grossly violated.

As a matter of fact, due to the large second-order effects,  $\pi^{\mu}_{\ \nu}$ is comparable in magnitude to $\epsilon$ in the above solutions. In other words, the Reynolds number is of order unity. This is clearly at the boundary of the region of validity of second-order theory (at least in the framework of Ref.~\cite{Denicol:2012cn}) and indicates the necessity of including even higher order terms. This also explains why its asymptotic behavior is different than that found in Navier-Stokes theory - the new solution found here is in a different regime of validity and it does not need to be smoothly connected to Navier-Stokes-like flow. Nevertheless, the existence of the nonperturbative scaling solutions of the type $\epsilon \sim 1/\tau^4$ demonstrated here may be preserved in higher-order theories since it follows from purely dimensional considerations and from the constraints for the transport coefficients set by conformal invariance.

\section{Conclusions}
\label{conc}
Building on our previous work \cite{Hatta:2014gqa}, we have constructed several new exact solutions of second-order hydrodynamic equations by conformally mapping Minkowski space to various curved spacetimes. A complicated flow in Minkowski space-time may look much simpler in another curved space-time, and this makes the systematic inclusion of second-order corrections possible when the flow has vanishing shear-tensor $\sigma^{\mu\nu}=0$. We have also learned that, by studying the same ideal solution in different curved spacetimes, one can obtain different set of second-order solutions.
 The case with nonvanishing $\sigma^{\mu\nu}$ is more difficult, but at least in one phenomenologically interesting case (Section \ref{shear}) we have been able to find special exact solutions to the most general second-order equations. These solutions may help to clarify the role played by each transport coefficient in second-order hydrodynamics.

 We emphasize that for our purposes it was crucial to treat $\pi^{\mu\nu}$ as independent variables rather than as being completely fixed by the gradient expansion. The former seems to be actually required in a consistent theory of relativistic hydrodynamics, namely, a theory that is stable and causal.
 When the hydrodynamics equations are analyzed nonperturbatively, either analytically or numerically, this makes a difference both qualitatively and quantitatively.
 In particular, in the examples considered in this paper, it is not permissible to use the `lowest-order' relation $\pi^{\mu\nu}\approx -2\eta \sigma^{\mu\nu}$ to treat $\pi^{\mu\nu}$ and $\sigma^{\mu\nu}$ interchangeably in the second-order terms.

 There are many directions for further study. Various other exact solutions may be obtained by considering a broader class of coordinate and Weyl transformations. For instance, it would be useful for phenomenological applications in heavy-ion collisions to find an extension of Gubser flow that is not radially symmetric in the transverse plane, which would allow one to investigate the role played by flow anisotropies in the hydrodynamic expansion of the quark-gluon plasma in an analytical manner. Finding more general solutions with $\sigma^{\mu\nu}\neq 0$ is also particularly challenging, nevertheless rather interesting. Moreover, the extension to nonconformal theories including the bulk pressure $\Pi$ is important in view of its potential impact in heavy-ion physics \cite{Noronha-Hostler:2013gga} and cosmology. Finally, it may also be interesting to explore the phenomenological relevance of the new boost-invariant solutions in the context of heavy-ion collisions.
   \\ \\

\noindent \textbf{Acknowledgements}

The authors thank the Yukawa Institute for Theoretical Physics, Kyoto University, where this collaboration started during the YITP-T-13-05 workshop ``New Frontiers in QCD". We thank  G.~S.~Denicol, Y.~Hidaka, and G.~Torrieri for discussions. J.~N. thanks Conselho Nacional de Desenvolvimento Cient\'ifico e Tecnol\'ogico (CNPq) and Funda\c c\~ao de Amparo \`a Pesquisa do
Estado de S\~ao Paulo (FAPESP) for financial support.

\end{document}